\documentclass[twocolumn,showpacs,amsmath,amssymb,aps,pra]{revtex4}
\usepackage{graphicx}
\usepackage{dcolumn}
\usepackage{bm}

\newcommand{\dd}{\displaystyle}

\begin{document}

\title{Atomic multipole relaxation rates near surfaces}

\author{J. A. Crosse}
\email{jac00@imperial.ac.uk}
\author{Stefan Scheel}
\email{s.scheel@imperial.ac.uk}
\affiliation{Quantum Optics and Laser Science, Blackett Laboratory,
Imperial College London, Prince Consort Road, London SW7 2AZ,
United Kingdom}

\date{\today}

\begin{abstract}
The spontaneous relaxation rates for an atom in free
space and close to an absorbing surface are calculated to various
orders of the electromagnetic multipole expansion. The spontaneous
decay rates for dipole, quadrupole and octupole transitions are
calculated in terms of their respective primitive electric multipole
moments and the magnetic relaxation rate is calculated for the dipole
and quadrupole transitions in terms of their respective primitive
magnetic multipole moments. The theory of electromagnetic field
quantization in magnetoelectric materials is used to derive general
expressions for the decay rates in terms of the dyadic Green
function. We focus on the decay rates in free space and near an
infinite half space. For the decay of atoms near to an absorbing
dielectric surface we find a hierarchy of scaling laws depending on
the atom-surface distance $z$.
\end{abstract}


\maketitle
\section{Introduction}
\label{sec:level1}

The spontaneous decay of an excited atom and the associated release of
a photon into the environment is due to the interaction between the atom and
the vacuum fluctuations of the electromagnetic field. It is a
fundamental process that is of critical importance in the
understanding of the dynamics of free atoms. It is, therefore,
unsurprising that this process has been the subject of much study,
some of which dates back almost to the beginnings of quantum mechanics
itself \cite{tolman}. The usual approach is to expand the atomic
charge distribution into a sum of its multipole moments. The
atom-field interaction is then considered to be dominated by the
linear coupling between the vacuum field and the dipole component of
the charge distribution. This approximation assumes the dipole-field
interaction to be sufficiently large that the higher-order multipole
terms can be neglected. Such atom-dipole calculations are a mainstay
of quantum optics textbooks
(e.g. \cite{cohenroc2,vogelwelsch,schubertwilhelmi,scully}). More
recently, extensive work has been done on dipole interactions in more
complex environments such as near absorbing surfaces
\cite{scheelknollwelsch2,dungknollwelsch2}, near an absorbing
microsphere \cite{microsphere,dungknollwelsch2} and in a spherical
microcavity \cite{dungknollwelsch2}.  

In the past few years, there has been increasing experimental interest
dipole-forbidden atomic transitions \cite{quad1,quad2}. In particular,
there has been a large body of work exploring the use of such higher
multipole moment transitions in atomic clocks, because the weak nature
of these transitions leads to a narrow line width and hence a much
better frequency standard \cite{gill}. Quadrupole transitions in 
$^{199}\mbox{Hg}^{+}$ \cite{hg,hg2} and $^{87/88}\mbox{Sr}^{+}$
\cite{sr,sr2,sr3} are frequently used and even the
$^{2}F_{\frac{7}{2}}-{}^{2}S_{\frac{1}{2}}$ octupole transitions
in $^{171}\mbox{Yb}^{+}$ has been studied \cite{yb}. Although there
have been some attempts to study the theory of quadrupole transitions
\cite{chance,klimov,klimov2,ducloy}, a complete understanding of the
nature of higher-order multipole decays is still lacking.

The corresponding magnetic interaction between the atom and the vacuum
fluctuations leads to a spontaneous change of the electronic spin
state. With the recent rise in interest in the magnetic trapping and
confinement of atoms for quantum control experiments and quantum
information processing \cite{magrev,magrev2,magrev3}, these spin flips
have become an increasingly important process. Atoms in specific
Zeeman sublevels of their hyperfine ground state can be magnetically
trapped. However, these states are subject to spin transitions which
can take the atom from a trapped state to an anti-trapped state
(i.e. the atom is actively expelled from the trap owing to the
relative orientation of the magnetic field and atomic spin
direction). This relaxation process is a limiting factor on the
lifetime of trapped atoms and may place serious restrictions on
experiments that require long trapping times such as those using a
trapped atom as a fundamental qubit for quantum information
processing. The dipole spin flip rate has been studied for a number of
different environments, for example above conducting and
superconducting surfaces
\cite{henkelpoettingwilkens,rekdal2,rekdal3,fermani} and close to
metal wires \cite{rekdal}.

In this article, we present a general theory of higher multipole
relaxation rates. The Green function method for the quantization of
the electromagnetic field is briefly discussed in Sec.~\ref{quant}. In
Sec.~\ref{mpc}, the multipole expansion and the multipole interaction
Hamiltonian are reviewed. In Sec.~\ref{eom}, the Heisenberg equation of
motion are used to derive general expressions for the decay rates for
various multipole orders from their respective interaction
Hamiltonians. Finally in Secs.~\ref{free} and \ref{surf}, the decay
of atoms in free space and close to a planar dielectric surface are
studied in detail. Some useful but lengthy calculations, in particular
on rotational averaging of tensors, can be found in the Appendices.

\section{Quantization scheme}
\label{quant}

The electromagnetic field in an absorbing magnetoelectric medium can be
quantized by expanding the electric field in terms of the dyadic Green
function (the inverse of the Helmholtz operator) and a set of bosonic
vector fields, which describe the collective excitations of the 
field and the absorbing matter (for reviews, see
e.g. \cite{scheelknollwelsch,slovaca}). The method begins with the
classical Maxwell equations in frequency space. In the absence of free
currents and charges these equations take the form
\begin{gather}
\bm{\nabla}\cdot\mathbf{B}(\mathbf{r},\omega) = 0,\\
\bm{\nabla}\times\mathbf{E}(\mathbf{r},\omega) -
i\omega\mathbf{B}(\mathbf{r},\omega) = \mathbf{0}, \label{faraday}\\
\bm{\nabla}\cdot\mathbf{D}(\mathbf{r},\omega) = 0,\\
\bm{\nabla}\times\mathbf{H}(\mathbf{r},\omega) +
i\omega\mathbf{D}(\mathbf{r},\omega) = \mathbf{0}. 
\end{gather}
Maxwell's equations have to be supplemented by constitutive relations
between the primary fields $\mathbf{E}(\mathbf{r},\omega)$ and
$\mathbf{B}(\mathbf{r},\omega)$, and the derived fields
$\mathbf{D}(\mathbf{r},\omega)$ and $\mathbf{H}(\mathbf{r},\omega)$,
respectively,
which
can be written as
\begin{gather}
\mathbf{D}(\mathbf{r},\omega) =
\varepsilon_0\mathbf{E}(\mathbf{r},\omega) +
\mathbf{P}(\mathbf{r},\omega) ,\\
\mathbf{H}(\mathbf{r},\omega) =
\kappa_0\mathbf{B}(\mathbf{r},\omega) -
\mathbf{M}(\mathbf{r},\omega)
\end{gather}
[$\kappa_0=1/\mu_0$].
The polarization field $\mathbf{P}(\mathbf{r},\omega)$ and the magnetization
field $\mathbf{M}(\mathbf{r},\omega)$ are, in the
linear-response approximation, related to the electric field and the 
magnetic induction by
\begin{gather}
\mathbf{P}(\mathbf{r},\omega) =
\varepsilon_0\left[\varepsilon(\mathbf{r},\omega)-1\right]
\mathbf{E}(\mathbf{r},\omega)+\mathbf{P}_\mathrm{N}(\mathbf{r},\omega),\\ 
\mathbf{M}(\mathbf{r},\omega) =
\kappa_0\left[1-\kappa(\mathbf{r},\omega)\right] 
\mathbf{B}(\mathbf{r},\omega) -\mathbf{M}_\mathrm{N}(\mathbf{r},\omega)
\end{gather}
where $\varepsilon(\mathbf{r},\omega)$ and
$\kappa(\mathbf{r},\omega)=1/\mu(\mathbf{r},\omega)$ are the
dielectric permittivity and the (inverse) magnetic permeability,
respectively.

Absorption in the medium is accounted for consistently by the addition of 
the noise polarization field $\mathbf{P}_\mathrm{N}(\mathbf{r},\omega)$ and the
noise magnetization field $\mathbf{M}_\mathrm{N}(\mathbf{r},\omega)$.
As a result of this, the (frequency components of the) electric field
obey the inhomogeneous Helmholtz equation
\begin{equation}
\bm{\nabla}\times\kappa(\mathbf{r},\omega)\bm{\nabla}\times
\mathbf{E}(\mathbf{r},\omega) -
\frac{\omega^2}{c^2}\varepsilon(\mathbf{r},\omega)
\mathbf{E}(\mathbf{r},\omega) 
= i\omega\mu_0 \mathbf{j}_\mathrm{N}(\mathbf{r},\omega)
\end{equation}
with the noise current density
\begin{equation}
\mathbf{j}_\mathrm{N}(\mathbf{r},\omega) =
-i\omega\mathbf{P}_\mathrm{N}(\mathbf{r},\omega) 
+\bm{\nabla}\times\mathbf{M}_\mathrm{N}(\mathbf{r},\omega) \,.
\end{equation}
This equation can be formally solved using the Green tensor for the
Helmholtz operator 
\begin{equation}
\mathbf{E}(\mathbf{r},\omega) = i\omega\mu_0\int d^3r'\,
\bm{G}(\mathbf{r},\mathbf{r}',\omega)\cdot
\mathbf{j}_\mathrm{N}(\mathbf{r}',\omega),
\end{equation}
where the Green tensor $\bm{G}(\mathbf{r},\mathbf{r}',\omega)$ has
the property 
\begin{equation}
\bm{\nabla}\times\kappa(\mathbf{r},\omega)\bm{\nabla}\times
\bm{G}(\mathbf{r},\mathbf{r}',\omega) - \frac{\omega^2}{c^2}
\varepsilon(\mathbf{r},\omega) \bm{G}(\mathbf{r},\mathbf{r}',\omega)
= \bm{\delta}(\mathbf{r}-\mathbf{r}').
\end{equation}

Quantization of the electromagnetic field is performed by decomposing
the noise polarization and magnetization fields in terms of two sets
of bosonic vector fields
\begin{gather}
\hat{\mathbf{P}}_\mathrm{N}(\mathbf{r},\omega) =
i\sqrt{\frac{\hbar\varepsilon_0}{\pi}\varepsilon''(\mathbf{r},\omega)}
\,\hat{\mathbf{f}}_e(\mathbf{r},\omega) ,\\
\hat{\mathbf{M}}_\mathrm{N}(\mathbf{r},\omega) =
i\sqrt{\frac{\hbar\kappa_0}{\pi}\kappa''(\mathbf{r},\omega)}
\,\hat{\mathbf{f}}_m(\mathbf{r},\omega) ,
\end{gather}
and imposing canonical commutation relations for them
[$\lambda,\lambda'=e,m$] 
\begin{equation}
\left[\hat{\mathbf{f}}_\lambda(\mathbf{r},\omega),
\hat{\mathbf{f}}^{\dagger}_{\lambda'}(\mathbf{r}',\omega')\right] =
\delta_{\lambda\lambda'} \bm{\delta}(\mathbf{r}-\mathbf{r}')
\delta(\omega-\omega'). 
\end{equation}
Thus the frequency components of the quantized electric 
field can be written as
\begin{equation}
\hat{\mathbf{E}}(\mathbf{r},\omega) = \sum\limits_{\lambda=e,m} \int
d^3r' \,\bm{G}_{\lambda}(\mathbf{r}, \mathbf{r}',
\omega)\cdot\hat{\mathbf{f}}_{\lambda}(\mathbf{r}',\omega), 
\label{E}
\end{equation}
with the magnetic induction field following from (\ref{faraday}),
and the abbreviations $\bm{G}_{\lambda}(\mathbf{r}, \mathbf{r}',
\omega)$ given by
\begin{align}
&\bm{G}_{e}(\mathbf{r}, \mathbf{r}', \omega) =
i\frac{\omega^{2}}{c^{2}}\sqrt{\frac{\hbar}{\pi\varepsilon_{0}}
\varepsilon''(\mathbf{r}',\omega)}
\bm{G}(\mathbf{r}, \mathbf{r}', \omega)\,,\\ 
&\bm{G}_{m}(\mathbf{r}, \mathbf{r}', \omega) =
-i\frac{\omega}{c}\sqrt{\frac{\hbar}{\pi\varepsilon_{0}}
\frac{\mu''(\mathbf{r}',\omega)}{|\mu(\mathbf{r}',\omega)|^2}}
\left[ \bm{G}(\mathbf{r},\mathbf{r}',\omega) \times
\overleftarrow{\bm{\nabla}}' \right].
\end{align}
The total electric-field operator reads
\begin{equation}
\hat{\mathbf{E}}(\mathbf{r}) = \int^{\infty}_{0}d\omega\,
\hat{\mathbf{E}}(\mathbf{r},\omega) + \mbox{h.c.}, \label{Etot}
\end{equation}
with a similar expression holding for the induction field.
The bosonic operators $\hat{\mathbf{f}}_\lambda(\mathbf{r}',\omega)$
and $\hat{\mathbf{f}}^{\dagger}_\lambda(\mathbf{r}',\omega)$ describe
collective excitations of the photonic modes and the absorbing
medium and can be viewed as the generalization of the free-space
photonic amplitude operators to arbitrary magnetoelectric  media. The
bilinear Hamiltonian
\begin{eqnarray}	
\label{HF}
\hat{H}_F = \sum\limits_{\lambda=e,m} \int d^3r\int d\omega\,
\hbar\omega\,
\hat{\mathbf{f}}^{\dagger}_\lambda(\mathbf{r},\omega)\cdot
\hat{\mathbf{f}}_\lambda(\mathbf{r},\omega)
\end{eqnarray}
can be used to generate the time-dependent Maxwell equations from
the Heisenberg equations of motion for the displacement field and the
magnetic field.

\section{Multipolar-coupling Hamiltonian and the
multipole expansion} 
\label{mpc}

The macroscopic interaction between light and matter is commonly
described using the multipolar coupling. In this picture, the matter is
described in terms of a polarization field, which is a result of
displaced charges within the material, and a magnetization field,
which is a result of charge currents within the material. The interaction 
terms take the form of linear couplings with
the polarization field interacting solely with the external electric
field and the magnetization field interacting solely with the external
magnetic field.

The Hamiltonian for a globally neutral system of point particles of
mass $m_{\alpha}$ and charge $q_{\alpha}$ in the presence of an
external electromagnetic field, whose centre of mass is at rest, 
has three contributions describing the medium-assisted quantized
electromagnetic field, the free motion of the charged particles and
the particle-field interaction, respectively \cite{slovaca,spin},
\begin{equation}
\hat{H} = \hat{H}_F + \hat{H}_A + \hat{H}_\mathrm{int} \,,
\label{ham}
\end{equation}
with
\begin{align}
\hat{H}_A &= \dd\sum_{\alpha}
\frac{\hat{\mathbf{p}}_{\alpha}^{2}}{2m_{a}} +
\frac{1}{2\varepsilon_{0}}\int d^{3}r\,\hat{\mathbf{P}}_A^2(\mathbf{r}),\\ 
\hat{H}_\mathrm{int} &= -\int d^{3}r\, \hat{\mathbf{P}}_A(\mathbf{r})\cdot
\hat{\mathbf{E}}(\mathbf{r}) - \int d^{3}r\,
\hat{\mathbf{M}}_A(\mathbf{r})\cdot\hat{\mathbf{B}}(\mathbf{r}),
\end{align}
and $\hat{H}_F$ given by Eq.~(\ref{HF}). Note here that diamagnetism,
which is quadratic in the magnetic field, is not
considered. Furthermore, since the centre of mass of the charge
distribution is assumed to be stationary, the
R\"{o}ntgen term vanishes. By considering the divergence of the
polarization field and the curl of the magnetization in terms of
displaced point charges and charge currents respectively, the
polarization and magnetization fields can be written in integral
representation as (see Appendix A)
\begin{equation}
\hat{\mathbf{P}}_A(\mathbf{r}) =
\dd\sum_{\alpha}q_{\alpha}\int_{0}^{1}ds\,\delta\left[\mathbf{r} -
\mathbf{r}_A -
s(\hat{\mathbf{r}}_{\alpha} - \mathbf{r}_A)\right]
(\hat{\mathbf{r}}_{\alpha} - \mathbf{r}_A)
\,,
\end{equation}
\begin{align}
\hat{\mathbf{M}}_A(\mathbf{r}) =
\dd\sum_{\alpha}\frac{q_{\alpha}}{2m_{\alpha}}&
\int_0^1ds\,s\,\delta\left[\mathbf{r} - \mathbf{r}_A -
s(\hat{\mathbf{r}}_{\alpha} - \mathbf{r}_A)\right]\nonumber\\
&\times\left[
(\hat{\mathbf{r}}_{\alpha}-\mathbf{r}_A)\times
\hat{\mathbf{p}}_{\alpha} - \hat{\mathbf{p}}_{\alpha}
\times(\hat{\mathbf{r}}_{\alpha}-\mathbf{r}_A)\right]\,\nonumber\\
& + \sum_{\alpha}\gamma_{\alpha}\hat{\mathbf{S}}_{\alpha}
\delta\left(\mathbf{r}-\hat{\mathbf{r}}_{\alpha}\right).  
\end{align}
Here $\hat{\mathbf{S}}_{\alpha}$ is the spin operator for particle
$\alpha$ located at $\hat{\mathbf{r}}_{\alpha}$,
$\hat{\mathbf{p}}_{\alpha}$ is the canonical momentum of the particle
relative to the atomic centre of mass and $\gamma_{\alpha}$ is the
gyromagnetic ratio of the particle (for an electron: $\gamma_{e} =
-eg_{e}/(2m_{e})$ with the electron $g$-factor $g_{e} \simeq
2$). Hence, the electric and magnetic terms in the interaction
Hamiltonian become 
\begin{align}
\hat{H}^{(E)}_\mathrm{int} =& -\int d^{3}r
\dd\sum_{\alpha}q_{\alpha}\int_{0}^{1}ds\,
\delta\left[\mathbf{r} -\mathbf{r}_A - s(\hat{\mathbf{r}}_{\alpha} -
\mathbf{r}_A)\right]
\nonumber\\ 
&\qquad\times(\hat{\mathbf{r}}_{\alpha} -
\mathbf{r}_A)\cdot\hat{\mathbf{E}}(\mathbf{r}), 
\label{he}
\end{align}
\begin{align}
\hat{H}^{(B)}_\mathrm{int} =& -\int d^{3}r
\dd\sum_{\alpha}\frac{q_{\alpha}}{2m_{\alpha}} \int_0^1ds
\,s\,\delta\left[\mathbf{r} - \mathbf{r}_A -
s(\hat{\mathbf{r}}_{\alpha} - \mathbf{r}_A)\right]\nonumber\\
&\times\left[
(\hat{\mathbf{r}}_{\alpha}-\mathbf{r}_A)\times
\hat{\mathbf{p}}_{\alpha} - \hat{\mathbf{p}}_{\alpha}
\times(\hat{\mathbf{r}}_{\alpha}-\mathbf{r}_A)
\right]\cdot\hat{\mathbf{B}}(\mathbf{r})\,\nonumber\\
& - \int d^{3}r
\dd\sum_{\alpha}\gamma_{\alpha}\hat{\mathbf{S}}_{\alpha}
\delta\left(\mathbf{r} - \hat{\mathbf{r}}_{\alpha}\right)
\cdot\hat{\mathbf{B}}(\mathbf{r}). 
\label{hb}
\end{align}

\subsection{Electric multipole expansion}
\label{sec:level2}

The interaction Hamiltonians (\ref{he}) and (\ref{hb}) are difficult
to handle and some simplifications are necessary to proceed. The usual
approach is to expand the polarization and magnetization fields in
terms of their multipole moments. The procedure is outlined here. More
information on the both the classical and quantum multipole expansion
can be found in Refs. \cite{jackson,raabdelange,cohenroc}. 

Consider first the electric term of the interaction Hamiltonian
(\ref{he}). The $\delta$-function can be expanded about the point
$\mathbf{r} = \mathbf{r}_A$ 
\begin{align}
\hat{H}^{(E)}_\mathrm{int} =& -\dd\sum_{\alpha}q_{\alpha}
\int d^3r\int_0^1ds\,
(\hat{\mathbf{r}}_{\alpha} -
\mathbf{r}_A)\cdot\hat{\mathbf{E}}(\mathbf{r})
\nonumber\\ 
&\times\bigg\{\delta(\mathbf{r} - \mathbf{r}_A) -
s(\hat{\mathbf{r}}_{\alpha} -
\mathbf{r}_A)\cdot\bm{\nabla}\delta(\mathbf{r} -
\mathbf{r}_A)
\nonumber\\ & \hspace*{-6ex}
+ \frac{s^{2}}{2!}[(\hat{\mathbf{r}}_{\alpha} -
\mathbf{r}_A)\cdot\bm{\nabla}]^{2}\delta(\mathbf{r} -
\mathbf{r}_A) +
O[(\hat{\mathbf{r}}_{\alpha}-\mathbf{r}_A)^3]\bigg\}.
\end{align}
Both integrals can now be performed. The spatial integral is evaluated
by integration by parts on the derivatives of the $\delta$-function.
As the vector $(\hat{\mathbf{r}}_{\alpha}-\mathbf{r}_A)$ is
not a function of the spatial variable $\mathbf{r}$, the derivatives
act only on the electric field. Thus the expansion becomes 
\begin{align}
&\hat{H}^{(E)}_\mathrm{int} =
-\dd\sum_{\alpha}q_{\alpha}(\hat{\mathbf{r}}_{\alpha} -
\mathbf{r}_A)\cdot\hat{\mathbf{E}}(\mathbf{r}_A) 
\nonumber\\
&\quad -\frac{1}{2!} \sum_{\alpha}q_{\alpha}[(\hat{\mathbf{r}}_{\alpha} -
\mathbf{r}_A)\otimes(\hat{\mathbf{r}}_{\alpha} -
\mathbf{r}_A)]\bullet[\bm{\nabla}\otimes\hat{\mathbf{E}}(\mathbf{r})]
|_{\mathbf{r}=\mathbf{r}_A}
\nonumber\\ 
&\quad -\frac{1}{3!}\dd\sum_{\alpha}q_{\alpha}[(\hat{\mathbf{r}}_{\alpha} -
\mathbf{r}_A)\otimes(\hat{\mathbf{r}}_{\alpha} -
\mathbf{r}_A)\otimes(\hat{\mathbf{r}}_{\alpha} -
\mathbf{r}_A)]
\nonumber\\
&\qquad\bullet[\bm{\nabla}\otimes\bm{\nabla}
\otimes\hat{\mathbf{E}}(\mathbf{r})]|_{\mathbf{r}
= \mathbf{r}_A} + \mathcal{O}\left[(\hat{\mathbf{r}}_{\alpha}
- \mathbf{r}_A)^{4}\right], 
\label{exp}
\end{align}
where the symbol $\bullet$ denotes the Hadamard product
($\mathbf{A}\bullet\mathbf{B}=A_{i_{1}\ldots i_{n}}B_{i_{1}\ldots i_{n}}$).
Equation~(\ref{exp}) can be written in a simpler form by defining
multipole moment (tensor) operators. The primitive electric dipole,
quadrupole and octupole moment operators are defined as 
\begin{align}
&\hat{\mathbf{d}} = \sum_{\alpha}q_{\alpha}(\hat{\mathbf{r}}_{\alpha}
- \mathbf{r}_A),\\ 
&\hat{\mathbf{d}}^{(4)} = \frac{1}{2!}
\dd\sum_{\alpha}q_{\alpha}(\hat{\mathbf{r}}_{\alpha} -
\mathbf{r}_A)\otimes(\hat{\mathbf{r}}_{\alpha} -
\mathbf{r}_A),\\
&\hat{\mathbf{d}}^{(8)} =
\frac{1}{3!}\dd\sum_{\alpha}q_{\alpha}(\hat{\mathbf{r}}_{\alpha} -
\mathbf{r}_A)\otimes(\hat{\mathbf{r}}_{\alpha} -
\mathbf{r}_A)\otimes(\hat{\mathbf{r}}_{\alpha} -
\mathbf{r}_A),
\end{align}
respectively. Note here that the multipole moments are defined as in
Refs.~\cite{schubertwilhelmi,cohenroc} with the coefficients of the
Taylor expansion absorbed into the definition of the moment. As a
result, the electric multipole interaction Hamiltonian can be written
as
\begin{align}
\hat{H}^{(E)}_\mathrm{int} =&
-\hat{\mathbf{d}}\cdot\hat{\mathbf{E}}(\mathbf{r}_A)
-\hat{\mathbf{d}}^{(4)}\bullet[\bm{\nabla}\otimes
\hat{\mathbf{E}}(\mathbf{r})]\mid_{\mathbf{r}
= \mathbf{r}_A}
\nonumber\\ &
-\hat{\mathbf{d}}^{(8)}\bullet[\bm{\nabla}\otimes\bm{\nabla}\otimes
\hat{\mathbf{E}}(\mathbf{r})]\mid_{\mathbf{r}
= \mathbf{r}_A}
+\mathcal{O}\left[(\hat{\mathbf{r}}_{\alpha}-\mathbf{r}_A)^4\right].
\label{exp2}
\end{align}

\subsection{Magnetic multipole expansion}

A similar expansion can be performed for the magnetic term of the
interaction Hamiltonian (\ref{hb}). Once again the $\delta$-function
can be expanded about the point $\mathbf{r} =
\mathbf{r}_A$. Performing the integrals as before and defining
the primitive magnetic dipole and quadrupole  moments as
\begin{align}
&\hat{\mathbf{m}} =
\frac{1}{2!}\sum_{\alpha}\left[q_{\alpha}(\hat{\mathbf{r}}_{\alpha} -
\mathbf{r}_A)\times\hat{\mathbf{p}}_{\alpha} +
2\gamma_{\alpha}\hat{\mathbf{S}}_{\alpha}\right]\\ 
&\hat{\mathbf{m}}^{(4)} =
\frac{1}{3!}\sum_{\alpha}\frac{q_{\alpha}}{m_{\alpha}}\bigg\{
\left[(\hat{\mathbf{r}}_{\alpha} -
\mathbf{r}_A)\times\hat{\mathbf{p}}_{\alpha}\right] \otimes
(\hat{\mathbf{r}}_{\alpha} - \mathbf{r}_A)\nonumber\\
& + \big((\hat{\mathbf{r}}_{\alpha} -
\mathbf{r}_A)\otimes\left[(\hat{\mathbf{r}}_{\alpha} -
\mathbf{r}_A)\times\hat{\mathbf{p}}_{\alpha}\right]\big)^{T}\bigg\}
\nonumber\\ &
+ \sum_{\alpha}(\hat{\mathbf{r}}_{\alpha} -
\mathbf{r}_A)\otimes\gamma_{\alpha}\hat{\mathbf{S}}_{\alpha}\,,
\end{align}
respectively, leads to a magnetic multipole interaction Hamiltonian of
the form 
\begin{align}
\hat{H}^{(B)}_\mathrm{int} =&
-\hat{\mathbf{m}}\cdot\hat{\mathbf{B}}(\mathbf{r}_A) -
\hat{\mathbf{m}}^{(4)}\bullet[\bm{\nabla}\otimes
\hat{\mathbf{B}}(\mathbf{r})] |_{\mathbf{r}=\mathbf{r}_A}
\nonumber\\
&\quad - \mathcal{O}\left[(\hat{\mathbf{r}}_{\alpha} -
\mathbf{r}_A)^3\right].
\label{exp3}
\end{align}

\section{Atomic equations of motion}
\label{eom}

The dynamics of atoms subject to electric dipole interaction with the
medium-assisted electromagnetic field has been studied extensively in
the literature
\cite{scheelknollwelsch2,dungknollwelsch2,microsphere}. More recently,
atoms subject to electric quadrupole interactions \cite{ducloy} and
magnetic dipole interactions \cite{rekdal,rekdal2,rekdal3,fermani}
have been considered. In order to analyse the effect of higher-order
multipole interactions on the atomic decay rates, we derive the
equation of motion for an atom subject to a single, specific 
higher-order interaction. We develop the Heisenberg equation of motion
for general multipoles and derive a general expression for multipole
driven relaxation rates. It is assumed that each transition is driven
by a single multipole interaction term and hence each multipole order
can be studied independently.  

Beginning with Eq.~(\ref{ham}), the microscopic multipolar-coupling
Hamiltonian can be written as
\begin{align}
\hat{H} = \hat{H}_F
+ \dd\sum_{i}\hbar\omega_{i}\hat{\sigma}_{ii} +
\hat{H}_\mathrm{int}.
\end{align}
Here $\hat{\sigma}_{ii}=|i\rangle\langle i|$ are the projection
operators  onto the energy eigenstates $|i\rangle$ with energies
$\hbar\omega_i$ of the atomic Hamiltonian, and $\hat{H}_F$ is given in
Eq.~(\ref{HF}). We define an operator $\hat{\mathcal{Q}^{\mathbf{r}}}$
that corresponds to the (differential) operator acting on the electric
field in the relevant term of the multipole expansion [e.g. for the
electric dipole $\hat{\mathcal{Q}}^{\mathbf{r}}=(\hat{\mathbf{d}}\cdot)$,
for the electric quadrupole $\hat{\mathcal{Q}}^{\mathbf{r}} =
(\hat{\mathbf{d}}^{(4)}\bullet\bm{\nabla}_{\mathbf{r}}\otimes)$, etc]. 
Furthermore, we can rewrite the interaction Hamiltonian as
\begin{align}
\hat{H}_\mathrm{int} &=
\hat{\mathcal{Q}}^{\mathbf{r}}\hat{\mathbf{E}}(\mathbf{r})\mid_{\mathbf{r}
= \mathbf{r}_{A}}\nonumber\\ 
&= -\left(\dd\sum_{i}\hat{\sigma}_{ii}\right)
\hat{\mathcal{Q}}^{\mathbf{r}}\left(\dd\sum_{j}\hat{\sigma}_{jj}\right)
\hat{\mathbf{E}}(\mathbf{r})\mid_{\mathbf{r}
= \mathbf{r}_{A}}\nonumber\\ 
&= -\sum_{ij}\mathcal{Q}_{ij}^{\mathbf{r}}\hat{\sigma}_{ij}
\hat{\mathbf{E}}(\mathbf{r})\mid_{\mathbf{r}
= \mathbf{r}_{A}} \,,
\end{align} 
where the electric field is given by Eqs.~(\ref{E}) and (\ref{Etot}),
and the completeness relation 
$\sum_{i}\hat{\sigma}_{ii} = \hat{1}$ has been used.
The Heisenberg equations of
motion for the atomic flip operators and the bosonic field operators
are thus
\begin{equation}
\dot{\hat{\mathbf{f}}}_{\lambda}(\mathbf{r},\omega) =
-i\omega\hat{\mathbf{f}}_{\lambda}(\mathbf{r},\omega) -
i\dd\sum_{ij}\bm{g}^\ast_{\lambda ,ij}(\mathbf{r}',\mathbf{r},\omega)
\hat{\sigma}_{ij}
\label{field}
\end{equation}
and
\begin{align}
\dot{\hat{\sigma}}_{ij} &=
i\omega_{ij}\hat{\sigma}_{ij}
- i\dd\sum\limits_{\lambda=e,m}\sum_{k}\int d^{3}s\int d\omega
\nonumber\\
&\left[\bm{g}_{\lambda ,jk}(\mathbf{r},\mathbf{s},\omega)
\hat{\sigma}_{ik}- \bm{g}_{\lambda ,ki}(\mathbf{r},\mathbf{s},\omega)
\hat{\sigma}_{kj}\right]
\cdot\hat{\mathbf{f}}_{\lambda}(\mathbf{s},\omega) + \mbox{h.c.}
\label{atom}
\end{align}
where the coupling tensors
$\bm{g}_{\lambda, ij}(\mathbf{r},\mathbf{r}',\omega)$ are defined by 
\begin{equation}
\bm{g}_{\lambda,ij}(\mathbf{r},\mathbf{r}',\omega) =
\frac{1}{i\hbar}\mathcal{Q}_{ij}^{\mathbf{r}} 
\bm{G}_{\lambda}(\mathbf{r},\mathbf{r}',\omega)
|_{\mathbf{r}= \mathbf{r}_{A}}\,.
\end{equation}

Our goal is to study the dynamics of the atom under the influence of
an external field, hence the next step is to remove the
electromagnetic degrees of freedom. This is done by solving the
equation of motion for the field (\ref{field}) and substituting the
solution into the equation of motion of the atom
(\ref{atom}). Equation~(\ref{field}) can be solved  
in the Markov approximation which assumes that atom to be memoryless
and hence not affected by previous field states.
Writing the atomic operator as a slowly varying envelope function and a
term oscillating with the atomic transition frequency,
$\hat{\sigma}_{ij}=\tilde{\hat{\sigma}}_{ij}e^{i\omega_{ij}t}$,
Eq.~(\ref{field}) can be integrated to give 
\begin{equation}
\hat{\mathbf{f}}_{\lambda}(\mathbf{r},\omega) = e^{-i\omega t}
\hat{\mathbf{f}}_{\lambda}^{\mathrm{free}}(\mathbf{r},\omega)
+ \dd\sum_{ij}\bm{g}^\ast_{\lambda ,ij}(\mathbf{r}',\mathbf{r},\omega)
\zeta(\omega-\omega_{ij})\hat{\sigma}_{ij}
\end{equation}
where $\zeta(\omega-\omega_{ij})$, in the long time limit, is defined as
\begin{equation}
\zeta(\omega - \omega_{ij}) = \mathcal{P}\frac{1}{(\omega -
\omega_{ij})} + i\pi\delta(\omega - \omega_{ij}) 
\end{equation}
with $\mathcal{P}$ denoting the principal value. Resubstituting the
expression for the field into the equation of motion for the atom
gives 
\begin{widetext}
\begin{align}
\dot{\hat{\sigma}}_{ij} &=
i\omega_{ij}\hat{\sigma}_{ij}
- i\dd\sum\limits_{\lambda=e,m}\sum_{k}\int d^{3}s\int d\omega
e^{-i\omega t} \left[ \bm{g}_{\lambda
,jk}(\mathbf{r},\mathbf{s},\omega)\hat{\sigma}_{ik} 
-\bm{g}_{\lambda ,ki}(\mathbf{r},\mathbf{s},\omega)
\hat{\sigma}_{kj} \right] \cdot
\hat{\mathbf{f}}_{\lambda}^{\mathrm{free}}(\mathbf{s},\omega)
\nonumber \\
&- i\dd\sum\limits_{\lambda=e,m}\sum_{k,p,q }\int d^{3}s\int d\omega
\left[ \bm{g}_{\lambda ,jk}(\mathbf{r},\mathbf{s},\omega)\hat{\sigma}_{ik}
-\bm{g}_{\lambda ,ki}(\mathbf{r},\mathbf{s},\omega)
\hat{\sigma}_{kj} \right] \cdot
\bm{g}^\ast_{\lambda, pq}(\mathbf{r}',\mathbf{s},\omega)\zeta(\omega
- \omega_{pq})\hat{\sigma}_{pq}.
\label{atomeom}
\end{align}
\end{widetext}
The first term in Eq.~(\ref{atomeom}) corresponds to the free motion
of the atom, the second to the interaction of the atom with the free
external field. The third term corresponds to an atom self
interaction mediated by the medium-assisted electromagnetic field. It
is this term that leads to spontaneous relaxation processes. 

State populations are described by the projection operator
$\hat{\sigma}_{ii}$. Since the free motion of a projection operator on
to a state is static the first term in Eq.~(\ref{atomeom})
vanishes. If there is no external field then the second term also
vanishes. The problem can be simplified further by studying only the
two relevant atomic levels that are involved in the transition. Since
the multipole moment matrix element contained in
$\mathcal{Q}_{ij}^{\mathbf{r}}$ vanishes for paired indices,
$\bm{g}_{ii}(\mathbf{r},\mathbf{s},\omega)$ vanishes and the evolution
of the excited state (here labelled as 2) can be reduced to
\begin{align}
\dot{\hat{\sigma}}_{22} &= i\sum\limits_{\lambda=e,m}\int d^{3}s\int
d\omega \nonumber\\ 
& \times \bigg[\bm{g}_{\lambda ,12}(\mathbf{r},\mathbf{s},\omega) \cdot
\bm{g}^\ast_{\lambda ,21}(\mathbf{r}',\mathbf{s},\omega)
\hat{\sigma}_{11}\zeta(\omega - \omega_{21})\nonumber\\
&- \bm{g}_{\lambda ,21}(\mathbf{r},\mathbf{s},\omega) \cdot
\bm{g}^\ast_{\lambda ,12}(\mathbf{r}',\mathbf{s},\omega)
\hat{\sigma}_{22}\zeta(\omega - \omega_{12}) \bigg].
\end{align}
The state labelled 1 indicates the ground state. Solving this equation
for the excited-state population gives 
\begin{align}
\hat{\sigma}_{22}(t) &= e^{-i(\tilde{\omega}-i\Gamma) t} \bigg[
\hat{\sigma}_{22}(0)
+ i\sum\limits_{\lambda=e,m}\int_0^t dt'\int d^{3}s\int d\omega
\nonumber\\ & \times
\bm{g}_{\lambda ,12}(\mathbf{r},\mathbf{s},\omega)  \cdot
\bm{g}^\ast_{\lambda ,21}(\mathbf{r}',\mathbf{s},\omega)
\hat{\sigma}_{11}(t')\zeta(\omega - \omega_{21}) \bigg] .
\end{align}
Here $\tilde{\omega} = \omega + \delta\omega$ is the shifted atomic transition frequency \cite{slovaca}. 
In the absence of an external field the level shift is small and hence can be neglected. 
The relaxation rate appears as an exponential decay factor in the free
evolution of the excited-state projection operator and is given by 
\begin{equation}
\Gamma = \pi \sum\limits_{\lambda=e,m}\int d^3s\,
\bm{g}_{\lambda ,21}(\mathbf{r},\mathbf{s},\omega_{21}) \cdot
\bm{g}^\ast_{\lambda ,12}(\mathbf{r}',\mathbf{s},\omega_{21}) .
\end{equation}
Expanding the coupling tensors
$\bm{g}_{\lambda,ij}(\mathbf{r},\mathbf{s},\omega)$
and using the integral relation for the Green functions,
\begin{multline}
\sum\limits_{\lambda=e,m}\int d^{3}s \,
\bm{G}_{\lambda}(\mathbf{r},\mathbf{s},\omega) \cdot
\bm{G}^{\ast}_{\lambda}(\mathbf{s},\mathbf{r}',\omega)   \\
=
\frac{\hbar\omega^{2}}{\pi c^{2}\varepsilon_{0}} 
\mathrm{Im}\,\bm{G}(\mathbf{r},\mathbf{r}',\omega) \,,
\end{multline}
the final expression for the decay rate is found to be 
\begin{equation}
\Gamma = \lim_{\mathbf{r},\mathbf{r}' \to
\mathbf{r}_{A}}\frac{\omega_{21}^{2}}{\hbar c^{2}\varepsilon_0}
\mathcal{Q}_{21}^{\mathbf{r}}
\mathrm{Im}\,\bm{G}(\mathbf{r},\mathbf{r}',\omega_{21})\mathcal{Q}_{12}^{\mathbf{r}'}.
\label{rate}
\end{equation}
Equation~(\ref{rate}) is a general result and hence true for all
multipole orders. In order to obtain any relaxation rate one only
needs to determine the form of the differential operator
$\mathcal{Q}_{ij}^{\mathbf{r}}$ for the relevant order, which itself
can be easily found from the multipole expansions (\ref{exp2}) and 
(\ref{exp3}) of the electromagnetic field. Together with the
definition of the multipole differential operator
$\mathcal{Q}_{ij}^{\mathbf{r}}$, Eq.~(\ref{rate}) constitutes
the main result of this article. One should note that Eq.~(\ref{rate})
  is consistant with the results of Ref. \cite{ducloy} 
where a general expression for the relaxation rate was derived using a classical 
approach. Substituting in the relevant expressions for 
$\mathcal{Q}_{ij}^{\mathbf{r}}$ the decay rates for various multipole
order can be found:
\begin{align}
&\Gamma =  \frac{\omega_{21}^{2}}{\hbar c^2\varepsilon_0}
\mathbf{d}_{21} \cdot  
\mathrm{Im}\,\bm{G}(\mathbf{r}_{A},\mathbf{r}_{A},\omega_{21})
\cdot\mathbf{d}_{12} \,, \label{spon2}\\
&\Gamma =  \lim_{\mathbf{r},\mathbf{r}' \to \mathbf{r}_{A}}
\frac{\omega_{21}^{2}}{\hbar c^{2}\varepsilon_{0}} \mathbf{d}^{(4)}_{21}
\bullet\bm{\nabla}\otimes 
\mathrm{Im}\,\bm{G}(\mathbf{r},\mathbf{r}',\omega_{21})\otimes
\overleftarrow{\bm{\nabla}}'\bullet\mathbf{d}^{(4)}_{12}
\,, \label{spon4}\\ 
&\Gamma =  \lim_{\mathbf{r},\mathbf{r}' \to \mathbf{r}_{A}}
\frac{\omega^{2}}{\hbar c^{2}\varepsilon_{0}}\times
\nonumber \\ & \quad
\mathbf{d}_{21}^{(8)}\bullet
\left(\bm{\nabla}
\otimes\bm{\nabla}\right)
\otimes \mathrm{Im}\,\bm{G}(\mathbf{r},\mathbf{r}',\omega_{21})
\otimes\left(\overleftarrow{\bm{\nabla}}'\otimes
\overleftarrow{\bm{\nabla}}'\right)\bullet
\mathbf{d}_{12}^{(8)} \,,\label{spon8}\\
&\Gamma =  \lim_{\mathbf{r},\mathbf{r}' \to
\mathbf{r}_{A}}\frac{\mu_{0}}{\hbar} 
\mathbf{m}_{21}\cdot\bm{\nabla}\times 
\mathrm{Im}\,\bm{G}(\mathbf{r},\mathbf{r}',\omega_{21})\times
\overleftarrow{\bm{\nabla}}'\cdot \mathbf{m}_{12}
\,,\label{flip2}\\ 
&\Gamma =  \lim_{\mathbf{r},\mathbf{r}' \to \mathbf{r}_{A}}
\frac{\mu_{0}}{\hbar}
\times\nonumber\\ & \quad
\mathbf{m}^{(4)}_{21}\bullet
\bm{\nabla}
\otimes\left[
\bm{\nabla}\times 
\mathrm{Im}\,\bm{G}(\mathbf{r},\mathbf{r}',\omega_{21})\times
\overleftarrow{\bm{\nabla}}'\right] 
\otimes\overleftarrow{\bm{\nabla}}'\bullet
\mathbf{m}^{(4)}_{12} \,.\label{flip4}
\end{align}
In these expressions, Eqs.~(\ref{spon2})---(\ref{spon8}) are the
electric dipole, quadrupole, and octupole transition rates,
respectively, and Eqs.~(\ref{flip2}) and (\ref{flip4}) the magnetic
dipole and quadrupole transition rates.

\section{Multipole decay rates in free space}
\label{free}

The expressions for the spontaneous decay and magnetic spin flip rates
in Eqs.~(\ref{spon2})---(\ref{flip4}) are valid for any environment
with the geometry of the system contained in the, as yet unspecified,
Green function. Calculating rates for different environments requires
knowledge of the specific Green function for that system. For
complicated geometries its calculation can be
highly involved, but in some simple cases it is
analytically known. In this section, the Green function for free space
is used to calculate the vacuum decay rates. In the next section, the
Green function for an infinite half space will be used to calculate
the decay rates of atoms near an absorbing dielectric surface. 

The Green function for free space, in component form, is given by
\begin{equation}
G^{(0)}_{\alpha\beta}(\mathbf{r},\mathbf{r}',\omega) =
\left(\partial^{\mathbf{r}}_{\alpha}\partial^{\mathbf{r}}_{\beta} 
+ q^2\delta_{\alpha\beta}\right)
\frac{e^{iq|\mathbf{r} - \mathbf{r}'|}}{4\pi q^2|\mathbf{r} - \mathbf{r}'|},
\label{freegreen}
\end{equation}
where $q=\omega/c$ is the wave number and the greek indices run over
the Cartesian coordinates $x$, $y$ and $z$.  

\subsubsection{Electric dipole decay rate}

Extensive work has been done on the dipole transition
\cite{scheelknollwelsch2,dungknollwelsch2,microsphere}. For
completeness the result is stated here. The spontaneous decay rate for
an electric dipole transition is given in component form by 
\begin{equation}
\label{eq:gamma_e2}
\Gamma = \lim_{\mathbf{r},\mathbf{r}' \to \mathbf{r}_{A}}
\frac{\omega^{2}}{\hbar c^{2}\varepsilon_{0}}d_{\alpha}d_{\beta}
\mathrm{Im}\,G_{\alpha\beta}(\mathbf{r},\mathbf{r}',\omega).
\end{equation}
Here the state indices have been dropped for clarity 
(hence $\omega_{12} = \omega_{21} = \omega$ and 
$\mathbf{d}_{12} = \mathbf{d}_{21} = \mathbf{d}$). From 
Eq.~(\ref{freegreen}), the imaginary part of the free space Green function 
in the coincidence limit $\mathbf{r}\to\mathbf{r}'$ is found to be
\begin{equation}
\mathrm{Im}\,G^{(0)}_{\alpha\beta}(\mathbf{r}_{A},\mathbf{r}_{A},\omega) = 
\frac{\omega}{6\pi c}\delta_{\alpha\beta}.
\end{equation}
Hence the free space dipole spontaneous decay rate is given by the
well known formula
\begin{equation}
\label{eq:gamma0_e2}
\Gamma_{0} = \frac{\omega^{3}}{6\pi\hbar c^{3}\varepsilon_{0}}
d_{\alpha}d_{\alpha}.
\end{equation}

\subsubsection{Electric quadrupole decay rate}

The quadrupole spontaneous decay rate in component form is given by
\begin{equation}
\label{eq:gamma_e4}
\Gamma =  \lim_{\mathbf{r},\mathbf{r}' \to \mathbf{r}_{A}}
\frac{\omega^{2}}{\hbar c^{2}\varepsilon_{0}}d^{(4)}_{\alpha\beta}
d^{(4)}_{\gamma\delta}\partial_{\alpha}
\partial '_{\gamma}
\mathrm{Im}\,G_{\beta\delta}(\mathbf{r},\mathbf{r}',\omega).
\end{equation}
Once again the state indices have been dropped for clarity. The
double derivatives of the imaginary part of the free space Green
function for coincident spatial variables are easily found. Note
that the both the Green function and $d^{(4)}_{\alpha\beta}$ are
symmetric in their respective indices. Using these symmetries it is
possible to show that the only non-vanishing terms are those with
paired indices in the outer product of the two quadrupole moment
vectors (sometimes referred to as the quadrupole polarizability
tensor). After collecting terms, the quadrupole spontaneous decay rate
in free space is found to be 
\begin{eqnarray}
\label{eq:gamma0_e4}
\Gamma_0 = \frac{\omega^{5}}{20\pi\hbar c^{5}\varepsilon_{0}}
\left\{d^{(4)}_{\alpha\beta}d^{(4)}_{\alpha\beta} -
\frac{1}{3}d^{(4)}_{\alpha\alpha}d^{(4)}_{\beta\beta}\right\}. 
\end{eqnarray}
Note here that the primitive moments have been used. In certain
circumstances (e.g. when the external field is divergence free), the
quadrupole moment tensor can be made traceless
\cite{jackson,raabdelange}. In these cases the second term in
Eq.~(\ref{eq:gamma0_e4}) vanishes. This returns the result which is
consistent with Ref.~\cite{ducloy} where the traceless quadrupole
moment tensor is used. However, traceless moments are not valid for
all physical situations (for example, if there are other sources in
the external field) and hence must be used with care. In contrast, the
primitive moments are always valid and hence the full form of
Eq.~(\ref{eq:gamma0_e4}) is true for all physical situations. 

\subsubsection{Electric octupole decay rate}

The octupole spontaneous decay rate in component form is given by
\begin{equation}
\label{eq:gamma_e8}
\Gamma =  \lim_{\mathbf{r},\mathbf{r}' \to
\mathbf{r}_{A}}\frac{\omega^{2}}{\hbar c^{2}\varepsilon_{0}}
d^{(8)}_{\alpha\beta\gamma} d^{(8)}_{\delta\eta\zeta}
\partial_{\alpha}\partial_{\beta}
\partial '_{\delta}\partial '_{\eta}
\mathrm{Im}\,G_{\gamma\zeta}(\mathbf{r},\mathbf{r}',\omega).
\end{equation}
The derivatives of the imaginary part of this function for coincident
spatial variables are again easily found. Here
$d^{(4)}_{\alpha\beta\gamma}$ is symmetric in all its indices. Thus,
again, all non-vanishing terms have paired indices in the octupole
polarizability tensor. Hence the octupole spontaneous decay rate in
free space can be written as
\begin{equation}
\label{eq:gamma0_e8}
\Gamma_0 = \frac{2\omega^{7}}{105\pi\hbar c^{7}\varepsilon_{0}}
\left\{d^{(8)}_{\alpha\beta\gamma}d^{(8)}_{\alpha\beta\gamma} -
\frac{1}{4}d^{(8)}_{\alpha\alpha\gamma}d^{(8)}_{\beta\beta\gamma}\right\}.
\end{equation}
As for the quadrupole decay rate, the primitive octupole moments have
been used. In certain circumstances trace constraints cause the second
term in Eq.~(\ref{eq:gamma0_e8}) to vanish. Note that for successive 
multipole orders the free space decay rates change by a factor proportional to 
$\omega^{2}R^{2}/c^{2}$ where $R$ is the characteristic length of the charge distribution of the atom.

\subsubsection{Magnetic dipole decay rate}

Let us now turn to magnetic transitions. The calculations are slightly
more involved due to the presence of the spatial derivatives acting
upon the Green tensor. In component form the magnetic dipole
transition rate is given by
\begin{equation}
\label{eq:gamma_m2}
\Gamma = \lim_{\mathbf{r},\mathbf{r}' \to \mathbf{r}_{A}}
\frac{\mu_{0}}{\hbar}m_{\alpha}m_{\mu}\varepsilon_{\alpha\beta\gamma}
\varepsilon_{\mu\nu\lambda}\partial _{\beta}
\partial '_{\nu}
\mathrm{Im}\,G_{\gamma\lambda}(\mathbf{r},\mathbf{r}',\omega).
\end{equation}
Calculating the derivatives one finds that the double curl of the
Green function in the coincidence limit is
\begin{equation}
\lim_{\mathbf{r},\mathbf{r}' \to \mathbf{r}_{A}}
\varepsilon_{\alpha\beta\gamma}\varepsilon_{\mu\nu\lambda}
\partial_{\beta}\partial '_{\nu}
\mathrm{Im}\,G^{(0)}_{\gamma\lambda}(\mathbf{r},\mathbf{r}',\omega)
=\frac{\omega^{3}}{6\pi c^{3}}\delta_{\alpha\beta}.
\end{equation}
Hence the magnetic dipole spin flip rate is given by
\begin{equation}
\label{eq:gamma0_m2}
\Gamma_{0} = \frac{\mu_{0}\omega^{3}}{6\pi\hbar c^{3}}m_{\alpha}m_{\alpha}.
\end{equation}
This is consistent with previous calculations based on dyadic Green
functions \cite{rekdal} as well as on employing Fermi's Golden Rule
\cite{purcell}.

\subsubsection{Magnetic quadrupole decay rate}

Finally, we calculate the quadrupole moment spin flip rate. The
notable complication at this order, compared to that of the electric
quadrupole moment, is that the primitive magnetic quadrupole moment is
not symmetric in its indices. In component form the quadrupole spin
flip rate is given by
\begin{multline}
\Gamma = \lim_{\mathbf{r},\mathbf{r}' \to \mathbf{r}_{A}}
\frac{\mu_{0}}{\hbar}m^{(4)}_{\alpha\beta}m^{(4)}_{\lambda\mu}
\varepsilon_{\beta\gamma\delta}\varepsilon_{\mu\nu\sigma}\\ \times
\partial_{\alpha}\partial_{\gamma}
\partial '_{\nu} \partial '_{\lambda}
\mathrm{Im}\,G_{\delta\sigma}(\mathbf{r},\mathbf{r}',\omega).
\label{eq:gamma_m4}
\end{multline}
Computing the derivatives one finds the magnetic quadrupole decay rate
in free space to be
\begin{equation}
\label{eq:gamma0_m4}
\Gamma_{0} = \frac{\mu_{0}\omega^{5}}{15\pi\hbar c^{5}}
\left\{m^{(4)}_{\alpha\beta}m^{(4)}_{\alpha\beta} -
\frac{1}{4}\left(m^{(4)}_{\alpha\beta}m^{(4)}_{\beta\alpha}
+m^{(4)}_{\alpha\alpha}m^{(4)}_{\beta\beta}\right)\right\}.
\end{equation}
As with the electric decay rates, successive magnetic 
multipole orders change by a factor proportional to 
$\omega^{2}R^{2}/c^{2}$.
From the above recipe, multipole decay rates of even higher order can
be easily derived, if needed.

\section{Multipole decay rates near absorbing surfaces}
\label{surf}

In this section the decay rates of atoms near an absorbing surface
are calculated. We shall consider surfaces that are not bianisotropic
(i.e. no cross-responses between electric and magnetic degrees of freedom).
Note also that diamagnetic materials are excluded from this consideration
as diamagnetism is an inherently nonlinear process.
Furthermore, we are interested in atoms that are subject to
either an electric or magnetic interaction. Thus there are in principle four
possible combinations of electric/magnetic atoms interacting with
electric/magnetic surfaces. However, the work involved can be reduced 
when one considers the duality properties of the electric and magnetic
fields. It was Silberstein \cite{duality1} who first suggested the
existence of symmetries in Maxwell's equations under exchange of the
electric and magnetic fields. This idea was further 
developed in Refs.~\cite{duality2,duality3}, and the concept has been
more formally established recently in Refs.~\cite{duality4,duality5} 
in the context of macroscopic QED. It has
been shown that under global exchange of the two fields various QED
quantities, in particular relaxation rates, remain unchanged.

Hence, in order to calculate the magnetic relaxation rate for an atom
above a purely magnetic surface one would only need to calculate the
electric relaxation rate for an atom near a purely electric surface
and apply the global transformations $\varepsilon \rightarrow \mu$ and
$\hat{\mathbf{d}} \rightarrow -\hat{\mathbf{m}}/c$. Similarly, in
order to find the electric relaxation rate of an atom above a purely
magnetic surface one only needs to calculate the magnetic relaxation
rate for an atom near a purely electric surface and then apply the
same global transformation. In light of this duality, we will only
calculate two of the four combinations and then deduce the other two
from duality. We shall formally calculate the interaction of atoms
above a purely electric surface since it is these systems, owing to
the dominant strength of electric interactions, that are of most
practical interest.

In the following, we envisage a medium that is infinitely extended
along the ($x,y$)-directions and layered in the $z$ direction such
that
\begin{align}
\varepsilon(\mathbf{r},\omega) = 
\left\{
\begin{array}{ll}
1,  & z>0\\
\varepsilon(\omega),  & z<0
\end{array}
\right. 
\end{align}
that is, an absorbing dielectric material of permittivity
$\varepsilon(\omega)$ occupying the lower half space with a
single interface to free space at $z=0$. The Green function can be
split into three parts,
\begin{align}
&\bm{G}(\mathbf{r},\mathbf{r}',\omega) \nonumber\\
&=\left\{
\begin{array}{ll}
\bm{G}^{bulk}(\mathbf{r},\mathbf{r}',\omega) +
\bm{R}(\mathbf{r},\mathbf{r}',\omega), & z,z'>0\;\mbox{or}\;z,z'<0 \\ 
\bm{T}(\mathbf{r},\mathbf{r}',\omega), & \mbox{otherwise}
\end{array}
\right.
\end{align}
where $\bm{T}(\mathbf{r},\mathbf{r}',\omega)$ is the Green
function for waves transmitted through the interface,
$\bm{R}(\mathbf{r},\mathbf{r}',\omega)$ is the Green function
for waves reflected at the interface and
$\bm{G}^{bulk}(\mathbf{r},\mathbf{r}',\omega)$ is the Green
function for a infinitely homogeneous, isotropic (bulk) medium of
permittivity equal to that of the layer where the spatial variables
are located. From the general expression for the decay rate
(\ref{rate}) it is clear that the Green function has to be evaluated
in the limit of coinciding spatial arguments. Hence, the transmission part
$\bm{T}(\mathbf{r},\mathbf{r}',\omega)$ does not play any role.
Furthermore, since the atom is located in free space above the
dielectric surface, the bulk contribution reduces to the free-space Green
function (\ref{freegreen}). Hence the decay rate of an atom close to
an absorbing surface can be decomposed into the free space decay rate
and a reflective correction owing to the presence of the surface 
\begin{equation}
\Gamma = \Gamma_{0} + \Gamma^{(S)}.
\label{total}
\end{equation}

The reflective part of the Green function can be written as 
the partial Fourier transform of its components 
\begin{equation}
\label{eq:weyl}
\bm{R}(\mathbf{r},\mathbf{r}',\omega) =
\int\frac{d^2k_\|}{(2\pi)^2}
\bm{R}(\mathbf{k}_\|,z,z'\omega)
e^{i\mathbf{k}_\|\cdot(\mathbf{r}_\|
- \mathbf{r}'_\|)} 
\end{equation}
where $\mathbf{r}_\| = (x,y,0)$ and $\mathbf{k}_\|=(k_{x},k_{y},0)$ are 
vectors restricted to the ($x,y$)-plane parallel to the dielectric
interface, and $k_\|=|\mathbf{k}_\||$. The Green tensor components for
a dielectric material in this specific geometry are listed in Appendix
B. The symbols $r^{TM}$ and $r^{TE}$ denote the Fresnel coefficients
for $TM$ and $TE$ waves
\begin{equation}
r^{TM} = \frac{\varepsilon(\omega)\beta_{z>0} -
\beta_{z<0}}{\varepsilon(\omega)\beta_{z>0} +
\beta_{z<0}},\hspace*{5mm}r^{TE} = \frac{\beta_{z>0} -
\beta_{z<0}}{\beta_{z>0} + \beta_{z<0}}, 
\end{equation}
with
\begin{equation}
\beta_{z<0} = \sqrt{q^{2}\varepsilon(\omega) -
k_\|^{2}},\hspace*{5mm}\beta_{z>0} = \sqrt{q^{2} -
k_\|^{2}}, 
\end{equation}
where $\mathrm{Im}\beta_{z \,\lessgtr\, 0}>0$ and $q=\omega/c$. The generic form of the Green function does not lend
itself to analytical investigations. However, in certain regimes it is 
possible to simplify the expression greatly. Here we consider the two 
limiting cases of near-field and far-field regimes. 

\subsection{Near field}

For a purely dielectric material the near-field approximation assumes
that the atom-surface distance is less than the effective transition wavelength, thus $z|\sqrt{\varepsilon(\omega)}|\omega/c \ll 1$ and
hence $q^{2},q^{2}|\varepsilon(\omega)| \ll k_\|^{2}$. Thus it is
possible to expand the expressions for $\beta_{z<0}$ and $\beta_{z>0}$
about $q^2/k_\|^2\varepsilon(\omega)$ and $q^2/k_\|^2$,
respectively. As a result,
\begin{align}
\beta_{z<0} &\approx ik_\|\left\{1 -
\frac{q^{2}}{2k_\|^{2}}\varepsilon(\omega) +
\mathcal{O}\left(\frac{q^{4}}{k_\|^{4}}
\varepsilon(\omega)^{2}\right)\right\},\nonumber\\
\beta_{z>0} &\approx ik_\|\left\{1 -
\frac{q^{2}}{2k_\|^{2}} +
\mathcal{O}\left(\frac{q^{4}}{k_\|^{4}}\right)\right\}. 
\end{align}
By considering only the leading order terms in these expansions, we
can find the leading order correction to the decay rates associated
with electric transitions. Thus, 
\begin{equation}
\beta_{z<0}\,, \beta_{z>0} \to ik_\|,\quad
r^{TM} \rightarrow \frac{\varepsilon(\omega)-1}{\varepsilon(\omega)+1},
\quad r^{TE} \rightarrow 0 ,
\end{equation}
and hence the reflective part of the Green function, in the near-field
approximation, to leading order is 
\begin{align}
&\bm{R}(\mathbf{k}_\|,z,z'\omega) \approx
\nonumber\\ 
&\frac{1}{2q^{2}}
\left[ \frac{\varepsilon(\omega)-1}{\varepsilon(\omega)+1} \right]
e^{-k_\|(|z| + |z'|)}
\left(\begin{array}{ccc} 
\frac{k_{x}^2}{k_\|} & \frac{k_{x}k_{y}}{k_\|} & -ik_{x} \\ 
\frac{k_{x}k_{y}}{k_\|} & \frac{k_{y}^2}{k_\|} & -ik_{y} \\ 
ik_{x} & ik_{y} & k_\|
\end{array}\right).
\label{LO}
\end{align}

For magnetic transitions near purely dielectric bodies, to leading order
in $\beta_{z<0}$ and $\beta_{z>0}$ there is no correction to the decay
rate. In this case, it is necessary to include the next-to-leading
order terms in $\beta_{z<0}$ and $\beta_{z>0}$,
\begin{equation}
\beta_{z<0} = ik_\| - i\frac{q^{2}}{2k_\|}
\varepsilon(\omega) \,,\quad
\beta_{z>0} = ik_\| - i\frac{q^{2}}{2k_\|}.
\end{equation}
Evaluating the double curl 
$\left[\bm{\nabla}\times\bm{R}(\mathbf{k}_\|,z,z'\omega)\times
\bm{\nabla}\right]$, expanding all occurrences of
$\beta_{z<0}$ and $\beta_{z>0}$ as above and neglecting terms of
order $\mathcal{O}\left(q^4/k_\|^2\right)$ one finds
that terms proportional to
$r^{TM}$ vanish. Only terms proportional to $r^{TE}$ remain which
itself can be approximated as
\begin{equation}
r^{TE} = \frac{\beta_{z>0} - \beta_{z<0}}{\beta_{z>0} +\beta_{z<0}}
\approx
\frac{q^{2}}{4k_\|^{2}}\left[\varepsilon(\omega) - 1\right].
\end{equation}
Hence, the double curl of the reflective part of the Green function,
in the near field approximation, to next-to-leading order is 
\begin{align}
&\left[\bm{\nabla}\times
\bm{R}(\mathbf{k}_\|,z,z'\omega)\times\bm{\nabla}\right] \approx\nonumber\\ 
&\frac{q^{2}}{8k_\|^{3}}\left[\varepsilon(\omega)-1\right]
e^{-k_\|(|z|+|z'|)}
\left(\begin{array}{ccc} 
k_{x}^2 & k_{x}k_{y} &
-ik_xk_\| \\
k_{x}k_{y} & k_{y}^{2} &
-ik_yk_\| \\
ik_xk_\| &
ik_yk_\| &
k_\|^2
\end{array}\right).
\label{NLO}
\end{align}
Equation~(\ref{LO}) will be used, together with
Eqs.~(\ref{spon2})---(\ref{spon8}), to find the transition rates
associated with electric transitions of an atom above a purely
dielectric surface and Eq.~(\ref{NLO}) will be used, together with 
Eqs.~(\ref{flip2}) and (\ref{flip4}), to calculate the corresponding
transition rates associated with magnetic transitions. 

\subsubsection{Electric multipole transitions}

In the following, we specify our general near-field results to the
electric dipole, quadrupole and octupole transitions. We start by
recalling the well-known results for electric-dipole transitions.
Upon using Eq.~(\ref{eq:gamma_e2}), the reflective correction to the
spontaneous dipole decay rate is found to be
\cite{surfdec1,surfdec2,surfdec3}
\begin{equation}
\Gamma^{(S)} = \frac{1}{16\pi\hbar\varepsilon_0z^3}
\frac{\varepsilon''(\omega)}{|\varepsilon(\omega)+1|^{2}}
\left\{d_{x}^{2} + d_{y}^{2} + 2d_{z}^{2}\right\}.
\label{nearsurfdipole}
\end{equation}
Although this is a general result, in many cases such detail is not
necessary. The above result assumes knowledge of the individual
components of the dipole moment vector, and hence the orientation of
the dipole moment.

For a free atom ensemble the dipole moment orientation of each
individual atom is generally unknown. For sufficiently large ensembles
of atoms, or for cases where the dipole orientation is unknown, 
it is possible to rotationally average the tensor formed by
the outer product of the two dipole moment vectors (also known as 
the polarizability tensor) over the sphere to obtain an
expression for the decay rate that is isotropic with respect to the
dipole moment orientation. The method used for the rotational
averaging involves contracting the polarizability tensor with an
averaging tensor (see Appendix C),
\begin{equation}
d^\mathrm{iso}_{i}d^\mathrm{iso}_{j} 
= I_{ij}^{\alpha\beta}d_{\alpha}d_{\beta}.
\end{equation}
Hence the spontaneous dipole decay rate becomes
\begin{equation}
\Gamma_\mathrm{iso}
= \frac{\omega^{2}}{\hbar c^{2}\varepsilon_{0}}
I_{ij}^{\alpha\beta}d_{\alpha}d_{\beta}
\mathrm{Im}\,G_{ij}(\mathbf{r}_{A},\mathbf{r}_{A},\omega).
\end{equation}
The rank-2 averaging tensor is
\begin{equation}
I_{ij}^{\alpha\beta} = \frac{1}{3}\delta_{\alpha\beta}\delta_{ij},
\end{equation}
which leads to a decay rate of
\begin{equation}
\Gamma_\mathrm{iso} = \frac{\omega^{2}}{3\hbar c^{2}\varepsilon_{0}}
d_{\alpha}d_{\alpha}\,
\mathrm{Im}\,G_{ii}(\mathbf{r}_{A},\mathbf{r}_{A},\omega).
\end{equation}
Note that the free-space decay rate is unchanged since the free-space
Green function for coincident spatial variables is already spherically
symmetric. The reflective correction to the decay rate thus becomes 
\begin{align}
\label{eq:gammanear_e2}
\Gamma_\mathrm{iso}^{(S)} &= \frac{1}{12\pi\hbar\varepsilon_0z^3}
\frac{\varepsilon''(\omega)}{|\varepsilon(\omega)+1|^{2}}
d_{\alpha}d_{\alpha}\nonumber\\ 
&= \frac{\varepsilon''(\omega)}{2|\varepsilon(\omega)+1|^{2}}
\left(\frac{c}{\omega z}\right)^{3} \Gamma_{0},
\end{align}
where $\Gamma_{0}$ is the dipole spontaneous decay rate 
(\ref{eq:gamma0_e2}) in free space.

Similar considerations lead to the decay rates associated with
higher-order electric multipole transitions.  The reflective
correction to the quadrupole spontaneous decay is given  by [recall
Eq.~(\ref{eq:gamma_e4})] 
\begin{align}
\Gamma^{(S)} =& \lim_{\mathbf{r},\mathbf{r}' \to\mathbf{r}_{A}}
\frac{\omega^{2}}{\hbar c^{2}\varepsilon_{0}}
d^{(4)}_{\alpha\beta}d^{(4)}_{\gamma\delta}\nonumber\\
&\times\partial_{\alpha}
\partial '_{\gamma}
\mathrm{Im}\left[\int\frac{d^2k_\|}{(2\pi)^2}
R_{\beta\delta}(\mathbf{k}_\|,z,z'\omega)
e^{i\mathbf{k}_\|\cdot(\mathbf{r}_\| - \mathbf{r}'_\|)}\right].
\end{align}
Remembering that the quadrupole moment tensor is symmetric in its 
indices gives a reflective correction to the quadrupole decay rate of 
\begin{equation}
\Gamma^{(S)} = \frac{3}{64\pi\hbar\varepsilon_0z^5}
\frac{\varepsilon''(\omega)}{|\varepsilon(\omega)+1|^{2}}
\left\{A_{\alpha\beta\gamma\delta}
d^{(4)}_{\alpha\beta}d^{(4)}_{\gamma\delta}\right\},
\end{equation}
where the term in curly brackets is a sum over the quadrupole moment
tensor components. The coefficients $A_{\alpha\beta\gamma\delta}$ for
the non-vanishing tensor components are given in Table~\ref{quadcoeff}.
\begin{table}[ht]
\centering                    
\begin{tabular}{c|c||c|c||c|c} 
\hline\hline                           
$d^{(4)}_{\alpha\beta}d^{(4)}_{\gamma\delta}$ & 
$A_{\alpha\beta\gamma\delta}$ &
$d^{(4)}_{\alpha\beta}d^{(4)}_{\gamma\delta}$ & 
$A_{\alpha\beta\gamma\delta}$ &
$d^{(4)}_{\alpha\beta}d^{(4)}_{\gamma\delta}$ &
$A_{\alpha\beta\gamma\delta}$ \\ 
\hline
$d^{(4)}_{xx}d^{(4)}_{xx}$ & 3 &
$d^{(4)}_{yy}d^{(4)}_{yy}$ & 3 &
$d^{(4)}_{zz}d^{(4)}_{zz}$ & 8  \\[1ex]  
$d^{(4)}_{xy}d^{(4)}_{xy}$ & 1 &
$d^{(4)}_{yz}d^{(4)}_{yz}$ & 4 &
$d^{(4)}_{zx}d^{(4)}_{zx}$ & 4  \\[1ex]   
$d^{(4)}_{xy}d^{(4)}_{yx}$ & 1 &
$d^{(4)}_{yz}d^{(4)}_{zy}$ & 4 &
$d^{(4)}_{zx}d^{(4)}_{xz}$ & 4  \\[1ex]   
$d^{(4)}_{yx}d^{(4)}_{xy}$ & 1 &
$d^{(4)}_{zy}d^{(4)}_{yz}$ & 4 &
$d^{(4)}_{xz}d^{(4)}_{zx}$ & 4  \\[1ex]   
$d^{(4)}_{yx}d^{(4)}_{yx}$ & 1 &
$d^{(4)}_{zy}d^{(4)}_{zy}$ & 4 &
$d^{(4)}_{zx}d^{(4)}_{zx}$ & 4  \\[1ex]   
$d^{(4)}_{xx}d^{(4)}_{yy}$ & 1 &
$d^{(4)}_{yy}d^{(4)}_{zz}$ & -4 &
$d^{(4)}_{zz}d^{(4)}_{xx}$ & -4  \\[1ex] 
$d^{(4)}_{yy}d^{(4)}_{xx}$ & 1 &
$d^{(4)}_{zz}d^{(4)}_{yy}$ & -4 &
$d^{(4)}_{xx}d^{(4)}_{zz}$ & -4  \\[1ex]   
\hline\hline         
\end{tabular}
\caption {Table of coefficients for the quadrupole spontaneous decay
rate of an atom close to an absorbing surface.} 
\label{quadcoeff}
\end{table}
We can again rotationally average over the quadrupole polarizability
tensor to obtain an isotropic expression for the decay rate of an
unorientated ensemble 
\begin{equation}
\Gamma_\mathrm{iso} = \lim_{\mathbf{r},\mathbf{r}' \to \mathbf{r}_{A}}
\frac{\omega^{2}}{\hbar c^{2}\varepsilon_{0}}
d^{(4)\mathrm{iso}}_{ij}d^{(4)\mathrm{iso}}_{km}\partial_{i}
\partial '_{k}
\mathrm{Im}\,G_{jm}(\mathbf{r},\mathbf{r}',\omega),
\end{equation}
with
\begin{equation}
d^{(4)\mathrm{iso}}_{ij}d^{(4)\mathrm{iso}}_{km} 
=I_{ijkm}^{\alpha\beta\gamma\delta}
d^{(4)}_{\alpha\beta}d^{(4)}_{\gamma\delta}.
\end{equation}
The full expression for $I_{ijkm}^{\alpha\beta\gamma\delta}$ is given
in Appendix C. 
We thus find a reflective correction to the spherically
averaged quadrupole decay rate of 
\begin{align}
\Gamma_\mathrm{iso}^{(S)} &= \frac{3}{10\pi\hbar\varepsilon_0z^5}
\frac{\varepsilon''(\omega)}{|\varepsilon(\omega)+1|^{2}}
\left\{d^{(4)}_{\alpha\beta}d^{(4)}_{\alpha\beta}
-\frac{1}{3}d^{(4)}_{\alpha\alpha}d^{(4)}_{\beta\beta}\right\}
\nonumber\\
&= 6\frac{\varepsilon''(\omega)}{|\varepsilon(\omega)+1|^{2}}
\left(\frac{c}{\omega z}\right)^{5} \Gamma_0,
\label{gammanear_e4}
\end{align}
where $\Gamma_{0}$ is the quadrupole spontaneous decay rate 
(\ref{eq:gamma0_e4}) in free space.

The same procedure can be repeated for the electric octupole transition
whose decay rate is given by [cf. Eq.~(\ref{eq:gamma_e8})]
\begin{align}
\Gamma^{(S)} =& \lim_{\mathbf{r},\mathbf{r}' \to\mathbf{r}_{A}}
\frac{\omega^{2}}{\hbar c^{2}\varepsilon_{0}}
d^{(8)}_{\alpha\beta\gamma}d^{(8)}_{\delta\eta\zeta}
\partial_{\alpha}\partial_{\beta}
\partial '_{\delta}\partial '_{\eta}
\nonumber\\
&\times
\mathrm{Im}\left[\int\frac{d^2k_\|}{(2\pi)^2}
R_{\gamma\zeta}(\mathbf{k}_\|,z,z'\omega)
e^{i\mathbf{k}_\|\cdot(\mathbf{r}_\| - \mathbf{r}'_\|)}\right].
\end{align}
By the same token, the
reflective correction to the octupole decay rate is
\begin{equation}
\Gamma^{(S)} = \frac{45}{256\pi\hbar\varepsilon_0z^7}
\frac{\varepsilon ''(\omega)}{|\varepsilon(\omega) + 1|^{2}}
\left\{A_{\alpha\beta\gamma\delta\mu\nu}
d^{(8)}_{\alpha\beta\gamma}d^{(8)}_{\delta\mu\nu}\right\},
\end{equation}
with coefficients $A_{\alpha\beta\gamma\delta\mu\nu}$ given in 
Table~\ref{octcoeff}. Note that the coefficients are unchanged under all
index permutations of the octupole moment tensor and by the
commutation of the two tensors. 
\begin{table}[ht]
\centering                        
\begin{tabular}{c|c||c|c} 
\hline\hline                           
$d^{(8)}_{\alpha\beta\gamma}d^{(8)}_{\delta\mu\nu}$ & 
$A^{(8)}_{\alpha\beta\gamma\delta\mu\nu}$ &
$d^{(8)}_{\alpha\beta\gamma}d^{(8)}_{\delta\mu\nu}$ & 
$A^{(8)}_{\alpha\beta\gamma\delta\mu\nu}$  \\ 
\hline
$d^{(8)}_{xxx}d^{(8)}_{xxx}$ & 5 & 
$d^{(8)}_{xzz}d^{(8)}_{xxx} + perm.$ & -6 \\[1ex] 
$d^{(8)}_{yyy}d^{(8)}_{yyy}$ & 5 &
$d^{(8)}_{yxx}d^{(8)}_{yyy} + perm.$ & 1 \\[1ex] 
$d^{(8)}_{zzz}d^{(8)}_{zzz}$ & 16 &
$d^{(8)}_{yzz}d^{(8)}_{yyy} + perm.$ & -6 \\[1ex]  
$d^{(8)}_{xxy}d^{(8)}_{xxy} + perm.$ & 1 &
$d^{(8)}_{zxx}d^{(8)}_{zzz} + perm.$ & -8 \\[1ex] 
$d^{(8)}_{xxz}d^{(8)}_{xxz} + perm.$ & 6 &
$d^{(8)}_{zyy}d^{(8)}_{zzz} + perm.$ & -8 \\[1ex]  
$d^{(8)}_{yyx}d^{(8)}_{yyx} + perm.$ & 1 &
$d^{(8)}_{xyy}d^{(8)}_{xzz} + perm.$ & -2 \\[1ex] 
$d^{(8)}_{yyz}d^{(8)}_{yyz} + perm.$ & 6 &
$d^{(8)}_{yzz}d^{(8)}_{yxx} + perm.$ & -2 \\[1ex]  
$d^{(8)}_{zzx}d^{(8)}_{zzx} + perm.$ & 8 &
$d^{(8)}_{zxx}d^{(8)}_{zyy} + perm.$ & 2 \\[1ex] 
$d^{(8)}_{zzy}d^{(8)}_{zzy} + perm.$ & 8 &
$d^{(8)}_{xyz}d^{(8)}_{xyz} + perm.$ & 2 \\[1ex]  
$d^{(8)}_{xyy}d^{(8)}_{xxx} + perm.$ & 1 & & \\[1ex] 
\hline\hline         
\end{tabular}
\caption {Table of coefficients for the octupole spontaneous decay
rate of an atom close to an absorbing surface.} 
\label{octcoeff}
\end{table}
For unorientated ensembles the decay rate becomes
\begin{equation}
\Gamma = \lim_{\mathbf{r},\mathbf{r}' \to \mathbf{r}_{A}}
\frac{\omega^{2}}{\hbar c^{2}\varepsilon_{0}}
d^{(8) iso}_{ijk}d^{(8) iso}_{pqr}
\partial _{i}\partial_{j}
\partial '_{p}\partial '_{q}
\mathrm{Im}\,G_{kr}(\mathbf{r},\mathbf{r}',\omega),
\end{equation}
with
\begin{equation}
d^{(8) iso}_{ijk}d^{(8) iso}_{pqr} =
I_{ijkpqr}^{\alpha\beta\gamma\delta\eta\zeta}
d^{(8)}_{\alpha\beta\gamma}d^{(8)}_{\delta\eta\zeta}.
\end{equation}
The full expression for
$I_{ijkpqr}^{\alpha\beta\gamma\delta\eta\zeta}$ is given in
Appendix~C. The 
reflective correction thus becomes 
\begin{align}
\Gamma_\mathrm{iso}^{(S)} &= \frac{9}{7\pi\hbar\varepsilon_0z^5}
\frac{\varepsilon''(\omega)}{|\varepsilon(\omega)+1|^{2}}
\left\{ d^{(8)}_{\alpha\beta\gamma}d^{(8)}_{\alpha\beta\gamma} -
\frac{1}{4}d^{(8)}_{\alpha\alpha\gamma}d^{(8)}_{\beta\beta\gamma}
\right\}
\nonumber\\
&=\frac{135}{2}\frac{\varepsilon''(\omega)}{|\varepsilon(\omega)+1|^{2}}
\left(\frac{c}{\omega z}\right)^{7} \Gamma_{0},
\label{gammanear_e8}
\end{align}
where $\Gamma_{0}$ is the octupole spontaneous decay rate 
(\ref{eq:gamma0_e8}) in free space.

Equations~(\ref{eq:gammanear_e2}), (\ref{gammanear_e4}) and 
(\ref{gammanear_e8}) reveal that with increasing order of the
electric multipole, the near-field scaling law of the relevant
spontaneous decay rates change by a factor proportional to 
$R^{2}/z^{2}$ where $R$ is the characteristic charge separation 
length of the atom and $z$ is the atom-surface distance.

\subsubsection{Magnetic multipole transitions}

We complete this study of near-field decay rates by investigating
magnetic multipole transitions. 
Magnetic dipole transitions in the presence of dielectric materials have
been intensively studied previously \cite{henkelpoettingwilkens,rekdal2}, 
and their results are presented here for completeness. The reflective 
correction to the magnetic dipole transition rate is given by 
[recall Eq.~(\ref{eq:gamma_m2})]
\begin{align}
&\Gamma^{(S)} = \lim_{\mathbf{r},\mathbf{r}' \to \mathbf{r}_{A}}
\frac{\mu_0}{\hbar}m_{\alpha}m_{\beta}
\nonumber\\ 
&\times \mathrm{Im}\left[\int\frac{d^2k_\|}{(2\pi)^2}
\left[\bm{\nabla}\times \bm{R}(\mathbf{k}_\|,z,z'\omega)\times
\bm{\nabla}\right]_{\alpha\beta}
\!e^{i\mathbf{k}_\|\cdot(\mathbf{r}_\|
- \mathbf{r}'_\|)}\right]
\end{align}
which, after inserting the expressions for the Green tensor, leads to
\begin{equation}
\Gamma^{(S)} = \frac{\mu_0\omega^2}{64\pi\hbar c^2z}
\varepsilon''(\omega)\{m_{x}^{2} + m_{y}^{2} + 2m_{z}^{2}\}.
\label{nearsurfdipolem}
\end{equation}
This is consistent with parts of the results in Refs.
\cite{rekdal2,henkelpoettingwilkens}, corresponding to spin flip rates 
for an atom above a metallic film in the case of large skin depth and 
film thickness. 

In the case of electric multipole transitions, it made sense to consider
unoriented atomic ensembles and to define rotationally averaged
spontaneous decay rates.
In principle the same averaging process could be performed
here. However, the result would not be physically meaningful since the
atomic spin is quantized about a specific quantization axis and
hence cannot have an arbitrary direction. Furthermore, in most
practical situations the spins of an ensemble of atoms are aligned by
external magnetic fields and hence assume a particular
orientation. Thus the rotationally averaged quantity is only of
academic interest and hence its calculation is renounced. 

Finally, the reflective contribution to the magnetic quadrupole decay 
rate is given by [recall Eq.~(\ref{eq:gamma_m4})]
\begin{align}
&\Gamma^{(S)} = \lim_{\mathbf{r},\mathbf{r}' \to \mathbf{r}_{A}}
\frac{\mu_0}{\hbar} m^{(4)}_{\alpha\beta}m^{(4)}_{\gamma\delta}
\partial_{\alpha}\partial '_{\gamma}
\nonumber\\
&\times \mathrm{Im}\left[\int\frac{d^2k_\|}{(2\pi)^2}
\left[\bm{\nabla}\times \bm{R}(\mathbf{k}_\|,z,z'\omega)\times
\bm{\nabla}\right]_{\beta\delta}
e^{i\mathbf{k}_\|\cdot(\mathbf{r}_\|
- \mathbf{r}'_\|)}\right]
\end{align}
which, in the near-field limit, can be written as
\begin{equation}
\Gamma^{(S)} = \frac{\mu_{0}\omega^2}{512\pi\hbar c^2z^3}
\varepsilon''(\omega)\{A_{\alpha\beta\gamma\delta}
m^{(4)}_{\alpha\beta}m^{(4)}_{\gamma\delta}\}.
\end{equation}
The coefficients for the non-vanishing components of the tensor 
$A_{\alpha\beta\gamma\delta}$ are given in Table~\ref{quadcoeff}
with the obvious interchange $d_{ij}\leftrightarrow m_{ij}$.
Note here that the coefficients are the same as those for the electric 
quadrupole decay rate. This is because they are properties of the geometry of the 
system and hence come from the Green function.
For the same reason
as given for the dipole case rotational averaging is not performed. As with 
the spontaneous decay rates, the magnetic transition rates also change 
by a factor proportional to $R^{2}/z^{2}$.

\subsection{Far field}

In the far-field limit one assumes that the distance to the surface
$z$ is large compared to the effective transition wavelength and thus $z|\sqrt{\varepsilon(\omega)}|\omega/c \gg 1$. In that
limit one can apply the method of stationary phase to compute the
first-order contribution to the Fourier integral (\ref{eq:weyl}). The
integral can be split into contributions from propagating waves
($k_\|\leq q$) and evanescent waves ($k_\|>q$), the latter of which
can be neglected in the far field. The propagating part consists of a
product of an amplitude function with a sinusoidal waveform whose
frequency is modulated by a further function with one stationary point
in the range of integration. At the stationary
point $k_\|=0$ we have
\begin{equation}
\beta_{z<0} = q\sqrt{\varepsilon(\omega)}\,,\quad
\beta_{z>0} = q,
\end{equation}
and therefore
\begin{equation}
r^{TE} \mapsto
\frac{1 - \sqrt{\varepsilon(\omega)}}{1 + \sqrt{\varepsilon(\omega)}} \,,\quad
r^{TM} \mapsto -r^{TE}.
\label{rte}
\end{equation}
Thus, in the far-field limit, the reflective part of the Green
function becomes
\begin{equation}
\mathbf{R}(\mathbf{k}_\|,z,z'\omega) =
\frac{i}{2q}
r^{TE}
e^{i\sqrt{q^{2} - k_\|^2}(|z| + |z'|)}
\mbox{diag}(1,1,0) \,.
\end{equation}
Analogously,
the double curl of the
reflective part of the Green function in the far-field limit is found
to be 
\begin{multline}
\left[\bm{\nabla}\times
\mathbf{R}(\mathbf{k}_\|,z,z'\omega)\times\bm{\nabla}\right] \\ 
=\frac{iq}{2} r^{TE}
e^{i\sqrt{q^{2} - k_\|^2}(|z| + |z'|)}
\mbox{diag}(1,1,0) \,.
\end{multline}

\subsubsection{Electric multipole transition rates}

The reflective corrections to the far-field electric dipole, quadrupole 
and octupole transition rates are given by Eqs.~(\ref{eq:gamma_e2}), 
(\ref{eq:gamma_e4}) and (\ref{eq:gamma_e8}), respectively.
By performing the inverse Fourier transform over $k_\|$ for each
component and taking the limit as 
$\mathbf{r},\mathbf{r}'\rightarrow\mathbf{r}_{A}$, the reflective 
correction to the spontaneous decay rates associated with dipole
transitions is found to be
\begin{equation}
\Gamma^{(S)} = 
\frac{\omega^{2}}{8\pi\hbar c^{2}\varepsilon_0z}\vert r^{TE}\vert
\sin \left( \frac{2\omega z}{c}+\arg r^{TE} \right)
\left\{d_{x}^{2} + d_{y}^{2}\right\}.
\end{equation}
Note here that $r^{TE}$ is the approximated coefficient given by
Eq.~(\ref{rte}). Similar considerations lead to the far-field
contributions to the decay rate associated with electric quadrupole
transitions,
\begin{multline}
\Gamma^{(S)} = - \frac{\omega^4}{8\pi\hbar c^4\varepsilon_0z} |r^{TE}|
\sin \left( \frac{2\omega z}{c}+\arg r^{TE} \right) \\ \times
\left\{d^{(4)}_{zx}d^{(4)}_{zx} + d^{(4)}_{zy}d^{(4)}_{zy}\right\}
\end{multline}
and octupole transitions, 
\begin{multline}
\Gamma^{(S)} = \frac{\omega^{6}}{8\pi\hbar c^{6}\varepsilon_0z} |r^{TE}|
\sin \left( \frac{2\omega z}{c}+\arg r^{TE} \right)
\\  \times
\left\{d^{(8)}_{zzx}d^{(8)}_{zzx} +
d^{(8)}_{zzy}d^{(8)}_{zzy}\right\},
\end{multline}
respectively.

As with the near-field decay rates these results can be rotationally
averaged using the appropriate rotational averaging tensors. For an
unorientated ensemble of atoms the $l$th electric multipole decay rate
becomes
\begin{equation}
\Gamma^{(S)}_\mathrm{iso} =  (-1)^{l+1}\frac{\Gamma_0}{2}\vert r^{TE}\vert
\sin \left( \frac{2\omega z}{c}+\arg r^{TE} \right)
\left(\frac{c}{\omega z}\right),
\end{equation}
where $\Gamma_0$ is the respective spontaneous decay rate in free space
(\ref{eq:gamma0_e2}), (\ref{eq:gamma0_e4}) or (\ref{eq:gamma0_e8}) for
dipole, quadrupole or octupole transitions, respectively. We see that successive
multipole orders change by a factor proportional to $\omega^{2}R^{2}/c^{2}$, 
just as the free space ones do.

\subsubsection{Magnetic multipole transition rates}

The reflective corrections to the transition rates associated with
magnetic dipole or quadrupole transitions is given by
Eqs.~(\ref{eq:gamma_m2}) and (\ref{eq:gamma_m4}), respectively.
By performing the inverse Fourier transform and taking the limit as
$\mathbf{r}, \mathbf{r} \rightarrow \mathbf{r}_{A}$, the reflective
correction to the dipole spin flip rate is found to be
\begin{equation}
\Gamma^{(S)} = -\frac{\mu_{0}\omega^{2}}{8\pi\hbar c^2z} |r^{TE}|
\sin \left( \frac{2\omega z}{c}+\arg r^{TE} \right)
\{m_x^2 + m_y^2\},
\end{equation}
and for the quadrupole rate one finds analogously
\begin{eqnarray}
\Gamma^{(S)} &=& \frac{\mu_{0}\omega^{4}}{8\pi\hbar c^4z} |r^{TE}|
\sin \left( \frac{2\omega z}{c}+\arg r^{TE} \right)
\nonumber \\ && \times
\left\{m^{(4)}_{zx}m^{(4)}_{zx} + m^{(4)}_{zy}m^{(4)}_{zy}\right\}.
\end{eqnarray}
As with the near-field transition rates the presence of a fixed
quantization axis prevents rotational averaging. Once again 
we see that successive multipole orders change by a factor 
proportional to $\omega^{2}R^{2}/c^{2}$.

In the far-field regime, both the electric and magnetic multipole
rates show a $1/z$-behaviour. This is to be expected as the far
fields created by all multipoles have the same distance scaling.

\subsection{Magnetic surfaces}

From the reflective correction for atoms above a dielectric surface,
the reflective correction for atoms above magnetically active surfaces
can be found using the duality of the electric and magnetic fields by
applying the transformation $\varepsilon \leftrightarrow \mu$ and
$\hat{\mathbf{d}} \rightarrow -\hat{\mathbf{m}}/c$ or
$\hat{\mathbf{m}} \rightarrow c\hat{\mathbf{d}}$. For example, the
near-field reflective correction to the dipole spin flip rate for an
atom above a magnetic surface can be found from
Eq.~(\ref{nearsurfdipole}) as
\begin{equation}
\Gamma^{(S)} = \frac{\mu_{0}}{16\pi\hbar z^3}
\frac{\mu''(\omega)}{|\mu(\omega)+1|^{2}}
\left\{m_{x}^{2} + m_{y}^{2} + 2m_{z}^{2}\right\}.
\end{equation}
Similarly, the near-field reflective correction to the dipole
spontaneous decay rate can be found from Eq.~(\ref{nearsurfdipolem})
as 
\begin{equation}
\Gamma^{(S)} = \frac{\omega^2}{64\pi\hbar\varepsilon_{0} c^2z}
\mu''(\omega)\{d_{x}^{2} + d_{y}^{2} + 2d_{z}^{2}\}.
\end{equation}
All other results convert in an identical fashion.

\section{Summary}
\label{sum}

Using the theory of electromagnetic field quantization in absorbing
magnetoelectric materials, the spontaneous decay and spin flip rates for the first
few multipole orders have been calculated. In the calculations the
primitive moments, as opposed to the traceless moments, have been used
since the primitive moments are applicable to all physical situations
whereas the traceless moments are only applicable where the external
field obeys the Laplace equation and hence is divergence free. The
decay rates for an atom in free space are given by
Eqs.~(\ref{eq:gamma0_e2}), (\ref{eq:gamma0_e4}), (\ref{eq:gamma0_e8}),
(\ref{eq:gamma0_m2}) and (\ref{eq:gamma0_m4}), respectively.

When considering an atom in the prescence of an absorbing surface, the
free-space decay rate is modified by a reflective correction which is
a result of wave reflection from the surface. In the near-field limit,
because of absorption by the surface, there are extra degrees of
freedom to which the atom can decay. The atom can lose energy
non-radiatively to the surface and hence the final state phase space
is larger and hence the decay rate is increased. The calculation shows
that the reflective correction is a complicated function of the
multipole orientation but there is also an inverse dependence on the
atom-surface distance $z$. The appearence of this type of scaling law 
is consistant with previous studies \cite{chance,klimov,klimov2,ducloy} 
where similar distance relations were found for the electric quadrupole 
decay rate in a variety of systems. The $z$-dependencies of the dominant order
in the reflective correction for the different multipole decay rates
are given in Table~\ref{order}.

\begin{table}[ht]
\centering                         
\begin{tabular}{c|c|c|c|c} 
\hline\hline
surface & \multicolumn{2}{c|}{electric} & \multicolumn{2}{c}{magnetic}\\
\hline
atom & electric & magnetic & electric & magnetic \\
\hline
dipole & $z^{-3}$ & $z^{-1}$ & $z^{-1}$ & $z^{-3}$\\
quadrupole & $z^{-5}$ & $z^{-3}$ & $z^{-3}$ & $z^{-5}$\\
octupole & $z^{-7}$ & & & $z^{-7}$ \\
\hline\hline         
\end{tabular}
\caption {Distance dependence of near-field multipole transition rates.}
\label{order}
\end{table}

Note that there is a clear hierarchy, with the power law dependence of
$z$ reducing by $2$ with each successive multipole order. It is clear
from the structure of the multipole expansion that this trend will
continue to higher order multipole moments. 

In the far field the reflective correction is inversely dependent on
the atom-surface distance to all orders of the multipole
expansion and hence the far fields created by all multipoles follow the same distance scaling law.
The other aspect of the far
field result is the sinusoidal nature of the reflective correction
with the atom-surface distance. This is caused by interference of
virtual electromagnetic waves reflected from the surface. 

\section{Acknowledgements}

This work was partially funded by the UK Engineering and Physical
Research Council. We would like to thank S.Y. Buhmann 
and R. Fermani for fruitful discussions.

\appendix

\section{s-integral parametrization of the polarization and
magnetization fields} 

A common parametrization for the polarization and magnetization fields
involves re-writing the polarization and magnetization fields as an
integral over and auxiliary parameter $s$. The derivations of these
parametrizations are briefly reviewed here. More information can be
found in Refs. \cite{vogelwelsch,cohenroc,babiker}. 

A collection of $\alpha$ point charges located at
$\mathbf{r}_{\alpha}$ can be described by the charge distribution
\begin{equation}
\hat{\rho}(\mathbf{r}) =
\dd\sum_{\alpha}q_{\alpha}\delta(\mathbf{r} -
\hat{\mathbf{r}}_{\alpha}) .
\label{dis} 
\end{equation}
In most cases, it is convenient to describe the system of charges by
a coarse-grained distribution
\begin{equation}
\rho_{A}(\mathbf{r}) = \left(\dd\sum_{\alpha}q_{\alpha}\right)
\delta(\mathbf{r} - \mathbf{r}_A).
\label{eqm}
\end{equation}
The difference between the actual charge distribution (\ref{dis}) and
its coarse-grained average defines the atomic polarization field via the 
implicit relation
\begin{equation}
\bm{\nabla}\cdot\hat{\mathbf{P}}_A(\mathbf{r}) =
-\hat{\rho}(\mathbf{r}) + \rho_{A}(\mathbf{r}). 
\label{pol}
\end{equation}
Fourier transforming with respect to the spatial variable
$\mathbf{r}$ gives 
\begin{equation}
i\mathbf{k}\cdot\tilde{\hat{\mathbf{P}}}_A(\mathbf{k}) =
\dd\sum_{\alpha}q_{\alpha}e^{-i\mathbf{k}\cdot\mathbf{r}_A}
\left[1 - e^{-i\mathbf{k}\cdot(\hat{\mathbf{r}}_{\alpha} -
\mathbf{r}_A)}\right] .
\end{equation}
This expression can be re-written in terms of an integral over an
auxiliary parameter $s$ 
\begin{equation}
\tilde{\hat{\mathbf{P}}}_A(\mathbf{k}) =
\dd\sum_{\alpha}q_{\alpha}e^{-i\mathbf{k}\cdot\mathbf{r}_A}
\int_{0}^{1}ds
(\hat{\mathbf{r}}_{\alpha} -
\mathbf{r}_A)e^{-i\mathbf{k}\cdot(\hat{\mathbf{r}}_{\alpha} -
\mathbf{r}_A)s}. 
\end{equation}
Performing the inverse Fourier transform gives the parametrized 
expression for the polarization field 
\begin{equation}
\hat{\mathbf{P}}_A(\mathbf{r}) =
\dd\sum_{\alpha}q_{\alpha}\int_{0}^{1}ds(\hat{\mathbf{r}}_{\alpha} -
\mathbf{r}_A)\delta\left[\mathbf{r} - \mathbf{r}_A -
s(\hat{\mathbf{r}}_{\alpha} - \mathbf{r}_A)\right]. 
\end{equation}

The magnetization field is generated by currents within the material
and hence is related to the rate of change of the distribution of
charges. Differentiating Eqn. (\ref{pol}) with respect to time and
applying the continuity equation 
$\dot{\hat{\rho}} +\bm{\nabla}\cdot\hat{\mathbf{j}} = 0$ gives 
\begin{equation}
\bm{\nabla}\cdot\left[\hat{\mathbf{j}}(\mathbf{r}) -
\hat{\mathbf{j}}_{A}(\mathbf{r}) -
\dot{\hat{\mathbf{P}}}_A(\mathbf{r})\right] = 0. 
\label{magcur}
\end{equation}
The quantity $\hat{\mathbf{j}}(\mathbf{r}) -\hat{\mathbf{j}}_{A}(\mathbf{r}) 
- \dot{\hat{\mathbf{P}}}_A(\mathbf{r})$
is known as the magnetization current. Since this represents
a divergence free quantity it can be written as the curl of another
vector field, the atomic magnetization $\hat{\mathbf{M}}_A(\mathbf{r})$ 
\begin{equation}
\bm{\nabla}\times\hat{\mathbf{M}}_A(\mathbf{r}) =
\hat{\mathbf{j}}(\mathbf{r}) -
\hat{\mathbf{j}}_{A}(\mathbf{r}) -
\dot{\hat{\mathbf{P}}}_A(\mathbf{r}). 
\end{equation}
For globally neutral ensembles of charges at rest,
$\hat{\mathbf{j}}_{A}(\mathbf{r})$
vanishes. The remaining two terms in Eqn.
(\ref{magcur}) are Fourier transformed with respect to
$\mathbf{r}$. Substituting in the parametrized expression for the
polarization field, performing the time differentiation and taking into 
account the spin contribution to $\hat{\mathbf{j}}(\mathbf{r})$ gives 
\begin{align}
& i\mathbf{k}\times\tilde{\hat{\mathbf{M}}}_A(\mathbf{k}) =
\dd\sum_{\alpha}q_{\alpha}\dot{\hat{\mathbf{r}}}_{\alpha}
e^{-i\mathbf{k}\cdot\hat{\mathbf{r}}_{\alpha}}
\nonumber\\
&\quad -\dd\sum_{\alpha}q_{\alpha}\int_{0}^{1}ds\dot{\hat{\mathbf{r}}}_{\alpha}
e^{-i\mathbf{k}\cdot\left[\mathbf{r}_A
+ s(\hat{\mathbf{r}}_{\alpha} 
- \mathbf{r}_A)\right]} 
\nonumber\\
&\quad -\dd\sum_{\alpha}q_{\alpha}\int_{0}^{1}ds
s(\hat{\mathbf{r}}_{\alpha}
-\mathbf{r}_A)(-i\mathbf{k}\cdot\dot{\hat{\mathbf{r}}}_{\alpha})
e^{-i\mathbf{k}\cdot\left[\mathbf{r}_A
+ s(\hat{\mathbf{r}}_{\alpha} -
\mathbf{r}_A)\right]}\nonumber\\
&\quad + \dd\sum_{\alpha}i\mathbf{k}\times\gamma_{\alpha}\hat{\mathbf{S}}e^{i\mathbf{k}\cdot\hat{\mathbf{r}}}. 
\end{align}
Integrating the second term on the rhs by parts and applying the vector 
identity
$\mathbf{A}\times(\mathbf{B}\times\mathbf{C}) =
\mathbf{B}(\mathbf{A}\cdot\mathbf{C}) -
\mathbf{C}(\mathbf{A}\cdot\mathbf{B})$
to the remaining terms gives
\begin{align}
\tilde{\hat{\mathbf{M}}}_A(\mathbf{k}) =&
\dd\sum_{\alpha}q_{\alpha}\int_{0}^{1}ds\,s
\left[(\hat{\mathbf{r}}_{\alpha}-
\mathbf{r}_A)\times\dot{\hat{\mathbf{r}}}_{\alpha}\right]
\nonumber\\ &\quad\times 
e^{-i\mathbf{k}\cdot\left[\mathbf{r}_A
+ s(\hat{\mathbf{r}}_{\alpha} -
\mathbf{r}_A)\right]}\nonumber\\
&\qquad + \dd\sum_{\alpha}\gamma_{\alpha}\hat{\mathbf{S}}e^{i\mathbf{k}\cdot\hat{\mathbf{r}}_{\alpha}}.
\end{align}
The final step
is to perform the inverse Fourier transform to give the parametrized
expression for the atomic magnetization field as
\begin{align}
\hat{\mathbf{M}}_A(\mathbf{r}) =&
\dd\sum_{\alpha}q_{\alpha}\int_{0}^{1}ds\,s\left[(\hat{\mathbf{r}}_{\alpha}
-\mathbf{r}_A)\times\dot{\hat{\mathbf{r}}}_{\alpha}\right]\nonumber\\
&\quad\times\delta\left[\mathbf{r} - \mathbf{r}_A -
s(\hat{\mathbf{r}}_{\alpha} - \mathbf{r}_A)\right] \nonumber\\
&\qquad + \sum_{\alpha}\gamma_{\alpha}\hat{\mathbf{S}}_{\alpha}\delta\left(\mathbf{r} - \hat{\mathbf{r}}_{\alpha}\right).
\end{align}
Note here that the physical variables have been used. For a Hamiltonian, the canonical momenta have to be introduced instead of the physical velocities. 

\section{Green function for a planar dielectric interface}

The reflective part of the Green function in terms of the partial
Fourier transform of its components is derived in Ref. \cite{tomas}, the
result of which is quoted here. Given that 
\begin{equation}
R_{\alpha\beta}(\mathbf{r},\mathbf{r}',\omega) =
\int\frac{d^2k_\|}{(2\pi)^2}
R_{\alpha\beta}(\mathbf{k}_\|,z,z'\omega)
e^{i\mathbf{k}_\|\cdot(\mathbf{r}_\|- \mathbf{r}'_\|)}, 
\end{equation}
with $\mathbf{r}_\| = (x,y,0)$ and $\mathbf{k}_\| = (k_{x},k_{y},0)$,
which are vectors restricted to the ($x,y$)-plane, and 
$k_\|=|\mathbf{k}_\||$, the components for the reflective part of the 
Green function for an infinitely extended planar dielectric material, 
which fills the lower half space $z<0$, are given by 
\begin{align}
&R_{xx} = \frac{-i}{2\beta_{z>0}}e^{i\beta_{z>0}(|z| + |z'|)}
\left[\frac{r^{TM}}{q^{2}}\beta_{z>0}^{2}\frac{k_{x}^{2}}{k_\|^{2}}
-r^{TE}\frac{k_{y}^{2}}{k_\|^{2}}\right],
\\
&R_{yy} = R_{xx}(k_x\leftrightarrow k_y)
\\
&R_{zz} = \frac{i}{2\beta_{z>0}}e^{i\beta_{z>0}(|z| + |z'|)}
\left[\frac{r^{TM}}{q^{2}}k_\|^{2}\right],
\\
&R_{xy} = R_{yx} \nonumber \\
&\quad =\frac{-i}{2\beta_{z>0}}e^{i\beta_{z>0}(|z| + |z'|)}
\left[\frac{r^{TM}}{q^{2}}\beta_{z>0}^{2}
\frac{k_{x}k_{y}}{k_\|^{2}} +r^{TE}\frac{k_{x}k_{y}}{k_\|^{2}}\right],
\\
&R_{xz} = \frac{-i}{2\beta_{z>0}}e^{i\beta_{z>0}(|z| + |z'|)}
\left[\frac{r^{TM}}{q^{2}}\beta_{z>0}k_{x}\right]
=-R_{zx},
\\
&R_{yz} = \frac{-i}{2\beta_{z>0}}e^{i\beta_{z>0}(|z| + |z'|)}
\left[\frac{r^{TM}}{q^{2}}\beta_{z>0}k_{y}\right]
=-R_{zy}.
\end{align}
The functions $r^{TM}$ and $r^{TE}$ are the Fresnel coefficients for 
$TM$ and $TE$ waves
\begin{equation}
r^{TM} = \frac{\varepsilon(\omega)\beta_{z>0}-\beta_{z<0}}
{\varepsilon(\omega)\beta_{z>0}+\beta_{z<0}},
\quad r^{TE} = \frac{\beta_{z>0}-\beta_{z<0}}{\beta_{z>0}+\beta_{z<0}}
\end{equation}
with
\begin{equation}
\beta_{z<0} = \sqrt{q^{2}\varepsilon(\omega) - k_\|^{2}},
\quad \beta_{z>0} = \sqrt{q^{2} - k_\|^{2}}
\end{equation}
and $q=\omega/c$.

\section{Rotational Averaging of Tensors}

In many cases physical systems are isotropic in nature but not
explicitly isotropic in their mathematical description. Substantial
simplification can be achieved by averaging of tensorial
quantities over the sphere. The description of the following method
follows closely the arguments in Ref. \cite{craig}. 

Contraction of a tensor with a rotation matrix whose components are
the cosines of an angle will rotate the coordinate system of the
tensor by that angle 
\begin{equation}
T_{i_{1}\cdots i_{n}} = I^{\alpha_{1}\cdots\alpha_{n}}_{i_{1}\cdots i_{n}}
T_{\alpha_{1}\cdots\alpha_{n}}.
\end{equation}
By integrating the rotation matrices over the Euler angles and
dividing by the spherical area the rotational average can be found 
\begin{equation}
\langle I^{\alpha_{1}\cdots\alpha_{n}}_{i_{1}\cdots i_{n}}\rangle =
\frac{1}{8\pi^{2}}\int_{0}^{2\pi} d\psi
\int_{0}^{2\pi} d\phi
\int_{0}^{\pi} \sin\theta\, d\theta\, 
I^{\alpha_{1}\cdots\alpha_{n}}_{i_{1}\cdots i_{n}},
\end{equation}
with
\begin{equation}
T^{iso}_{i_{1}\cdots i_{n}} = \langle 
I^{\alpha_{1}\cdots\alpha_{n}}_{i_{1}\cdots i_{n}}\rangle 
T_{\alpha_{1}\cdots\alpha_{n}}.
\end{equation}
This is cumbersome even for tensors of low rank. A simpler method is
to contract the tensor with an `averaging tensor'.  Owing to its
spherical symmetry, any averaging tensor of rank $n$ can be written
as a sum of the basic linearly independent rank-$n$ isomers which in
turn can be constructed from the two fundamental isotropic tensors in
three dimensions; the Kronecker delta $\delta_{ij}$ and the
Levi-Civita symbol $\varepsilon_{ijk}$. Consider a Cartesian frame of
reference in three dimensions $\mathbb{V}$ with a set of basic rank-$n$ 
isomers $\bm{f}^{(n)}$ where the initial non-isotropic tensor
lies and a Cartesian frame of reference $\mathbb{U}$ with a set of
basic rank-$n$ isomers $\bm{g}^{(n)}$ in which the rotationally
averaged tensor lies. The averaging tensor $\bm{I}^{(n)}$ is a
rank-$2n$ tensor that maps between the two spaces,
$\bm{I}:\mathbb{V}\rightarrow\mathbb{U}$, such that 
\begin{equation}
\bm{g}^{n} = \bm{I}^{(n)}\cdot\bm{f}^{n}.
\label{isomer}
\end{equation}
Furthermore,
\begin{equation}
\bm{I}^{(n)} = \bm{g}^{n}\cdot\mathbf{M}\cdot\bm{f}^{n},
\label{aveten}
\end{equation}
where $\mathbf{M}$ is a matrix of coefficients. Since the isomers for each
rank are known all that is required is to determine $\mathbf{I}^{(n)}$
is to find $\mathbf{M}$. Combining Eqs.~(\ref{isomer}) and (\ref{aveten})
yields
\begin{eqnarray}
\bm{g}^{n} &=&  \bm{g}^{n}\cdot\mathbf{M}\cdot
\left(\bm{f}^{n}\cdot\bm{f}^{n}\right),\nonumber\\
\left(\bm{g}^{n}\cdot\bm{g}^{n}\right) &=&  
\left(\bm{g}^{n}\cdot\bm{g}^{n}\right)\cdot\mathbf{M}\cdot
\left(\bm{f}^{n}\cdot\bm{f}^{n}\right).
\end{eqnarray}
Note that the coordinates of $\mathbb{V}$ and
$\mathbb{U}$ are both Cartesian (merely rotated with respect to each
other), hence the isomers for the two spaces have the same form and
magnitude 
\begin{equation}
\left(\bm{g}^{n}\cdot\bm{g}^{n}\right) 
= \left(\bm{f}^{n}\cdot\bm{f}^{n}\right) = \mathbf{S}^{(n)},
\end{equation}
where $\mathbf{S}^{(n)}$ is a matrix formed from all possible
contractions of each isomer in the set. Thus 
\begin{equation}
\mathbf{S}^{(n)} =  \mathbf{S}^{(n)}\cdot\mathbf{M}\cdot\mathbf{S}^{(n)}
\Rightarrow
\mathbf{M} = \left(\mathbf{S}^{(n)}\right)^{-1}.
\end{equation}
Hence, $\mathbf{I}^{(n)}$ can be determined purely from the isomers. 

The rank-2 averaging tensor, used in the rotational averaging of
the dipole decay rate, is constructed from the single
rank-2 isomer
$f^{(2)} = \delta_{\alpha\beta}$,
hence $\mathbf{S}^{(2)}$ is a single number
\begin{equation}
\mathbf{S}^{(2)} = f^{(2)}f^{(2)} 
= \delta_{\alpha\beta}\delta_{\alpha\beta} 
= 3,
\end{equation}
$\mathbf{M}=1/3$
and thus
\begin{equation}
I_{ij}^{\alpha\beta} = \frac{1}{3} \delta_{ij}\delta_{\alpha\beta}.
\end{equation}

As a non-trivial example we calculate the averaging tensor for a
rank-4 tensor used in the rotational averaging of the quadrupole
decay rate. The rank-4 isomers are 
\begin{equation}
f_{1}^{(4)} = \delta_{\alpha\beta}\delta_{\mu\nu},\quad
f_{2}^{(4)} = \delta_{\alpha\mu}\delta_{\beta\nu},\quad
f_{3}^{(4)} = \delta_{\alpha\nu}\delta_{\beta\mu}.
\end{equation}
The elements of $\mathbf{S}^{(4)}$ are
\begin{equation}
S^{(4)}_{ij} = f^{(4)}_{i}f^{(4)}_{j},
\end{equation}
so for example
\begin{align}
&S^{(4)}_{11} = f^{(4)}_{1}f^{(4)}_{1} 
= \delta_{\alpha\beta}\delta_{\mu\nu}
\delta_{\alpha\beta}\delta_{\mu\nu} 
= 9,
\nonumber\\
&S^{(4)}_{12} = f^{(4)}_{1}f^{(4)}_{2} 
= \delta_{\alpha\beta}\delta_{\mu\nu} 
\delta_{\alpha\mu}\delta_{\beta\nu} 
= 3.
\end{align}
Hence,
\begin{equation}
\mathbf{S}^{(4)} = 
\left(\begin{array}{ccc} 
9 & 3 & 3 \\ 3 & 9 & 3 \\ 3 & 3 & 9 
\end{array}\right) 
\Rightarrow
\mathbf{M}= \frac{1}{30}
\left(\begin{array}{ccc} 
4 & -1 & -1 \\ -1 & 4 & -1 \\ -1 & -1 & 4 
\end{array}\right) ,
\end{equation}
so that we can write the rank-4 averaging tensor as
\begin{equation}
I_{ijkm}^{\alpha\beta\mu\nu} = \frac{1}{30}
\left(\begin{array}{ccc} 
\delta_{ij}\delta_{km} \\ \delta_{ik}\delta_{jm} \\ 
\delta_{im}\delta_{jk} \end{array}\right) ^{T}
\left(\begin{array}{ccc} 
4 & -1 & -1 \\ -1 & 4 & -1 \\ -1 & -1 & 4 
\end{array}\right) 
\left(\begin{array}{ccc} 
\delta_{\alpha\beta}\delta_{\mu\nu} \\ 
\delta_{\alpha\mu}\delta_{\beta\nu} \\  
\delta_{\alpha\nu}\delta_{\beta\mu} 
\end{array}\right).
\end{equation}

Finally, using the above method, the rank-6 rotational averaging
tensor, used in the calculation of the octupole spontaneous decay
rate, is given below.
\begin{widetext}
\begin{equation}
I_{ijkpqr}^{\alpha\beta\gamma\delta\eta\zeta} = \frac{1}{210}
\left(\begin{array}{c} 
\delta_{ij}\delta_{kp}\delta_{qr} \\ 
\delta_{ij}\delta_{kq}\delta_{pr} \\ 
\delta_{ij}\delta_{kr}\delta_{pq} \\ 
\delta_{ik}\delta_{jp}\delta_{qr} \\ 
\delta_{ik}\delta_{jq}\delta_{pr} \\ 
\delta_{ik}\delta_{jr}\delta_{pq} \\ 
\delta_{ip}\delta_{jk}\delta_{qr} \\ 
\delta_{ip}\delta_{jq}\delta_{kr} \\ 
\delta_{ip}\delta_{jr}\delta_{kq} \\ 
\delta_{iq}\delta_{jk}\delta_{pr} \\ 
\delta_{iq}\delta_{jp}\delta_{kr} \\ 
\delta_{iq}\delta_{jr}\delta_{kp} \\ 
\delta_{ir}\delta_{jk}\delta_{pq} \\ 
\delta_{ir}\delta_{jp}\delta_{kq} \\ 
\delta_{ir}\delta_{jq}\delta_{kp} \\ 
 \end{array}\right)^{T}
\left(\begin{array}{ccccccccccccccc} 
16 & -5 & -5 & -5 & 2 & 2 & -5 & 2 & 2 & 2 & 2 & -5 & 2 & 2 & -5 \\
-5 & 16 & -5 & 2 & -5 & 2 & 2 & 2 & -5 & -5 & 2 & 2 & 2 & -5 & 2 \\
-5 & -5 & 16 & 2 & 2 & -5 & 2 & -5 & 2 & 2 & -5 & 2 & -5 & 2 & 2 \\
-5 & 2 & 2 & 16 & -5 & -5 & -5 & 2 & 2 & 2 & -5 & 2 & 2 & -5 & 2 \\
2 & -5 & 2 & -5 & 16 & -5 & 2 & -5 & 2 & -5 & 2 & 2 & 2 & 2 & -5 \\
2 & 2 & -5 & -5 & -5 & 16 & 2 & 2 & -5 & 2 & 2 & -5 & -5 & 2 & 2 \\
-5 & 2 & 2 & -5 & 2 & 2 & 16 & -5 & -5 & -5 & 2 & 2 & -5 & 2 & 2 \\
2 & 2 & -5 & 2 & -5 & 2 & -5 & 16 & -5 & 2 & -5 & 2 & 2 & 2 & -5 \\
2 & -5 & 2 & 2 & 2 & -5 & -5 & -5 & 16 & 2 & 2 & -5 & 2 & -5 & 2 \\
2 & -5 & 2 & 2 & -5 & 2 & -5 & 2 & 2 & 16 & -5 & -5 & -5 & 2 & 2 \\
2 & 2 & -5 & -5 & 2 & 2 & 2 & -5 & 2 & -5 & 16 & -5 & 2 & -5 & 2 \\
-5 & 2 & 2 & 2 & 2 & -5 & 2 & 2 & -5 & -5 & -5 & 16 & 2 & 2 & -5 \\
2 & 2 & -5 & 2 & 2 & -5 & -5 & 2 & 2 & -5 & 2 & 2 & 16 & -5 & -5 \\
2 & -5 & 2 & -5 & 2 & 2 & 2 & 2 & -5 & 2 & -5 & 2 & -5 & 16 & -5 \\
-5 & 2 & 2 & 2 & -5 & 2 & 2 & -5 & 2 & 2 & 2 & -5 & -5 & -5 & 16 \\
\end{array}\right)
\left(\begin{array}{c} 
\delta_{\alpha\beta}\delta_{\gamma\delta}\delta_{\eta\zeta} \\ 
\delta_{\alpha\beta}\delta_{\gamma\eta}\delta_{\delta\zeta} \\ 
\delta_{\alpha\beta}\delta_{\gamma\zeta}\delta_{\delta\eta} \\ 
\delta_{\alpha\gamma}\delta_{\beta\delta}\delta_{\eta\zeta} \\ 
\delta_{\alpha\gamma}\delta_{\beta\eta}\delta_{\delta\zeta} \\ 
\delta_{\alpha\gamma}\delta_{\beta\zeta}\delta_{\delta\eta} \\ 
\delta_{\alpha\delta}\delta_{\beta\gamma}\delta_{\eta\zeta} \\ 
\delta_{\alpha\delta}\delta_{\beta\eta}\delta_{\gamma\zeta} \\ 
\delta_{\alpha\delta}\delta_{\beta\zeta}\delta_{\gamma\eta} \\ 
\delta_{\alpha\eta}\delta_{\beta\gamma}\delta_{\delta\zeta} \\ 
\delta_{\alpha\eta}\delta_{\beta\delta}\delta_{\gamma\zeta} \\ 
\delta_{\alpha\eta}\delta_{\beta\zeta}\delta_{\gamma\delta} \\ 
\delta_{\alpha\zeta}\delta_{\beta\gamma}\delta_{\delta\eta} \\ 
\delta_{\alpha\zeta}\delta_{\beta\delta}\delta_{\gamma\eta} \\ 
\delta_{\alpha\zeta}\delta_{\beta\eta}\delta_{\gamma\delta} \\ 
 \end{array}\right).
\end{equation}
\end{widetext}


\end{document}